\documentclass{tlp}
\usepackage{alltt}
\usepackage{amsmath}
\usepackage{CJKutf8}
\usepackage[dvipsnames]{xcolor} 
\usepackage{enumitem}
\usepackage{float} 
\usepackage{graphicx}
\usepackage{url}
\usepackage{xspace}
\usepackage[noend,ruled,linesnumbered]{algorithm2e} 
\usepackage{epsfig}
\usepackage{hyperref} 

\newcommand{\longline}{\noindent\rule{\textwidth}{.01in}}
\newcommand{\ErgoAI}{\ensuremath{\cal E}rgoAI\xspace}
\newcommand{\hilog}{HiLog\xspace}
\newcommand{\Ergo}{\ensuremath{\cal E}rgoAI\xspace}
\newcommand{\Flora}{{\em ${\cal F}$lora-2}\xspace}
\newcommand{\cP}{{\cal P}}
\newcommand{\naf}{\texttt{$\backslash$naf}\xspace}
\newcommand{\eneg}{\texttt{$\backslash$neg}\xspace}
\newcommand{\opposes}{\texttt{$\backslash$opposes}\xspace}
\newcommand{\cancel}{\texttt{$\backslash$cancel}\xspace}
\newcommand{\overrides}{\texttt{$\backslash$overrides}\xspace}
\newcommand{\dateTime}{\texttt{$\backslash$dateTime}}
\newcommand{\isa}{\texttt{:}\xspace}
\newcommand{\rawisa}{\texttt{:}\xspace}
\newcommand{\mysub}{\texttt{::}\xspace}

\newcommand{\plg}{\textbackslash plg\xspace}

\newcommand{\mvd}{\texttt{->}\xspace}
\newcommand{\bs}{\textbackslash}

\newcommand{\saltapply}{\texttt{'}\textnormal{\textit{salt}\texttt{\textasciicircum}}\texttt{apply}\texttt{'}\xspace}
\newcommand{\saltmvd}{\texttt{'}\textnormal{\textit{salt}\texttt{\textasciicircum}}\texttt{mvd}\texttt{'}\xspace}
\newcommand{\saltisa}{\texttt{'}\textnormal{\textit{salt}\texttt{\textasciicircum}}\texttt{isa}\texttt{'}\xspace}
\newcommand{\saltapplymod}{\texttt{'}\textnormal{\textit{salt}\texttt{\textasciicircum}\textit{modname}\texttt{\textasciicircum}}\texttt{apply}\texttt{'}\xspace}
\newcommand{\saltisamod}{\texttt{'}\textnormal{\textit{salt}\texttt{\textasciicircum}\textit{modname}\texttt{\textasciicircum}}\texttt{isa}\texttt{'}\xspace}
\newcommand{\saltmvdmod}{\texttt{'}\textnormal{\textit{salt}\texttt{\textasciicircum}\textit{modname}\texttt{\textasciicircum}}\texttt{mvd}\texttt{'}\xspace}

\newcommand{\integer}{\texttt{$\backslash$integer}\xspace}
\newcommand{\typecheck}{\texttt{$\backslash$typecheck}\xspace}

\newcommand{\true}{{\em true}\xspace}
\newcommand{\false}{{\em false}\xspace}
\newcommand{\undefined}{{\em u}\xspace}
\newcommand{\SLGI}{{\em SLGR}\xspace}
\newcommand{\SLGIA}{{\em SLGR(A)}\xspace}
\newcommand{\SLGRA}{{\em SLGR(A)}\xspace}
\newcommand{\revision}{\color{black}\xspace}

\newtheorem{definition}{Definition} 
\newtheorem{example}{Example} 

\begin{document}

\title[Multi-paradigm Logic Programming in the ErgoAI System]
      {Multi-paradigm Logic Programming\\ in the \ErgoAI System}

\author[Michael Kifer and Theresa Swift]
       {MICHAEL KIFER\thanks{Supported in part by the NSF grant 1814457}   \\
         Stony Brook University, USA, and
         Coherent Knowledge, USA \\
         \email{kifer@cs.stonybrook.edu}
         \and THERESA SWIFT \\
         Coherent Knowledge, USA \\
         \email{Theresasturn@gmail.com}}

       \maketitle

       \begin{abstract}
  \ErgoAI is a high-level, multi-paradigm logic programming language
  and system developed by Coherent Knowledge Systems as an enhancement
  of and a successor to the popular
  \Flora
  system.  \Ergo is oriented towards scalable
  knowledge representation and reasoning, and can exploit 
  structured knowledge as well as knowledge derived from external sources
  such as vector databases.  From their inception, \Ergo (and \Flora before
  it) were designed to exploit the well-founded semantics for
  reasoning in a multi-paradigm environment, including
  \emph{object-based} logic (F-logic) with non-monotonic inheritance;
  \emph{higher order} syntax in the style of HiLog;
  \emph{defeasibility} of rules; semantically clean 
  \emph{transactional updates};
  \emph{subgoal delay} for handling unsafe queries and
  for better performance; and optional support for bounded rationality
  at a module level.
  Although \Ergo programs are compiled into XSB
  and adopt many Prolog features, \Ergo is altogether a different
  language and system.



%

Under consideration in Theory and Practice of Logic Programming (TPLP).
\end{abstract}



       \section{Introduction}\label{sec:intro}

\ErgoAI is an open-source knowledge representation and reasoning (KRR)
system oriented towards scalable reasoning about hybrid knowledge in a
dynamically changing environment.  \Ergo is the successor to and
replacement for the well-known \Flora KRR
system~\cite{flora2-odbase-2003,flora2site}, which is no longer
maintained.  \Ergo includes many features either not in, or much
improved from those in \Flora.\footnote{This paper is a significant
extension of \cite{SwiK24}.}

The language of \Ergo is based on \hilog (\emph{higher-order} syntax)
\cite{CKW93} and Frame logic (abbr. F-logic) supporting objects,
types, inheritance \cite{KiLW95} set within a strong dynamic module
system.  In addition, \Ergo cleanly integrates general formulas with
the \emph{exists} and \emph{forall} quantifiers; user-defined
functions;
\emph{defeasible} reasoning in the
LPDA style \cite{WGKFL09};
\emph{transactional updates}~\cite{BoKi94} with integrity constraints
and alerts; 
bounded rationality~\cite{GroS13,RigS14}; and dynamic subgoal
reordering.  \Ergo has been designed so that these extensions fit 
together hand in glove, allowing for a high degree of coherence both in
language and semantically.

\Ergo treats applications written in other languages,
such as Prolog, Python, and C, simply as special modules, which enables
tight integration.  For instance, \Ergo can use knowledge
available in Python, such as vector embeddings, in combination with
traditionally structured knowledge.  Due to space limitations, this
paper covers only some of the aforesaid paradigms.

Unlike KRR systems, such as ASP solvers and description logic provers,
\Ergo is based on the Well-Founded Semantics (WFS) \cite{VRS91} which
has low-order polynomial data complexity.  By choosing a weaker logic,
\Ergo gains several advantages.  For instance, \Ergo evaluations do not need to be
grounded: as will be shown, even non-ground negation is soundly
evaluated.  The lack of need for grounding allows \Ergo to use
unrestricted logical terms
as data structures---a use that distinguishes \Ergo from Datalog
systems.  However \Ergo shares with Datalog (and Prolog) the
ability to evaluate queries in a relevant, top-down manner.

Although \Ergo programs are compiled into XSB Prolog \cite{SwiW12},
\Ergo relies on its own reader, optimizing compiler, and runtime
libraries.~\footnote{A rough measure of the sophistication of \Ergo is
that it is based on over 200,000 lines of Prolog, \Ergo, and C code,
while its manuals have a length of over 600 pages.} At the same time,
\Ergo relies heavily on XSB's tabled evaluation and directly benefits
from XSB's close connection with Python through the Janus
interface~\cite{SwiA23}.

\Ergo demonstrates that a variety of new logical features can be
incorporated into a fast, robust, and scalable system. For instance,
in \Ergo all predicates and frames support \hilog by default, so that
the functor of a term or predicate can be any (not necessarily ground)
term, as opposed to
an atomic functor, as in Prolog.  Similarly, although Prologs such as
XSB, SWI, Ciao, and YAP
  support tabled resolution, they use
SLDNF resolution as a default.  In \Ergo the {\em default} resolution
method is SLG resolution for WFS, extended with ({\em R})eactive incremental
tabling, and with optional bounded rationality through subgoal and
answer ({\em A})bstraction, a resolution method we term \SLGIA.  If desired
though, a predicate $P$ can be declared as non-tabled if $P$ contains
side-effects, or for other purposes.

The journey from Flora-2 to \ErgoAI began in 2013 when a small group
formed Coherent Knowledge Systems to develop Flora-2 into a more
robust, commercially supported system.  \Ergo extends Flora-2 in
having an IDE, explanation capabilities, full reactive tabling,
instantiation delay, termination support, and tripwires, as well as
many smaller features.  However, perhaps the most effort was spent to
make \Ergo more robust and scalable both in \Ergo's runtime libraries
and at the XSB level.
As a result of \ErgoAI's features and stability, the system is being used
for research and commercial applications in software assurance
\cite{Bloom24}, financial compliance, tax accounting and legal
reasoning.\footnote{The website for Coherent Knowledge Systems
\url{http://coherentknowledge.com} lists a number of applications in
the above areas. Also, see the TED talk on legal reasoning which
contains a discussion of legal modeling with \Ergo~\cite{Morris18}}. In
addition it is currently being used in the DARPA CODORD project as a
target language for the generation of executable logic from English
text \cite{CODORDweb}.\footnote{\ErgoAI is open-source and available at
{\tt https://github.com/ErgoAI}.}

This paper describes several of \ErgoAI's logical extensions at a
technical but accessible level. The first group of {\em non-monotonic}
extensions
includes defeasible inheritance, the use of default and explicit
negation, and argumentation theories that are adjoined to programs to
provide different flavors of defeasibility.  Non-monotonic features make use
of the truth value \undefined, which represents truth values that are
either \emph{undefined} in WFS or that are \emph{unknown} due to the
use of bounded rationality.  The second group of {\em dynamic}
capabilities
includes {\em reactive tabling} that automatically updates inferences
when data changes,
along with
integrity constraints that signal and take actions whenever updated
inferences fulfill specified conditions.  These dynamic capabilities
work within the framework of {\em transaction logic}
to ensure that each update is semantically meaningful.
A third {\em query evaluation} group (1) 
minimizes floundering of unsafe goals (including negation) through
{\em instantiation delay}; and (2) supports sound termination of {\em any} 
program through a type of bounded rationality called {\em restraint}.

The paper is structured as follows.  Section~\ref{sec:prelim} provides
background on WFS and \SLGIA, \ErgoAI's tabled resolution method.
Sections~\ref{sec:frames} and \ref{sec:hilog} describe the base
language of \Ergo: frame-logic with inheritance and
\hilog.\footnote{This language is sometimes termed {\em
  Rulelog}~\cite{ABCDFGKLS13}.}  Section~\ref{sec:modules} overviews
\Ergo's module system, which allows code files to be associated with
modules in a completely dynamic manner.  Section~\ref{sec:defeas}
describes \ErgoAI's default argumentation theory for defeasible
reasoning and Section~\ref{sec:flexsearch} its approach to soundly
evaluate unsafe goals.  Section~\ref{sec:dynamic} discusses \ErgoAI's
approach to KRR in a dynamically changing environment, while
Section~\ref{sec:brat} discusses \ErgoAI's approaches to termination
including sound termination through bounded rationality.
Section~\ref{sec:architecture} provides a summary of \Ergo's
architecture.  Finally, Section~\ref{sec:perf} provides results on
scalability along with comparisons of \ErgoAI's speed and memory
consumption to that of Prolog.

When reading the following sections that describe \Ergo features
(Sections~\ref{sec:prelim}-\ref{sec:brat}) users may wish to refer to 
Section~\ref{sec:architecture}, which provides a high-level summary
and diagram of \Ergo's architecture.


       \section{Preliminaries: The Well-Founded Semantics and Tabling}\label{sec:prelim}

In the late 1980s the semantics of negation in logic programming was
an active topic of research
from which two main semantics emerged: the stable model
semantics~\cite{GeLi88} and WFS~\cite{VRS91}.  These semantics are
related, as the well-founded model of a program $P$ is contained
within the intersection of every stable model of $P$~\cite{Przy89c}
and so is a total stable model in the case of programs with
stratified negation, and a partial stable model otherwise.
The stable model semantics has given rise to the important field of Answer
Set Programming (cf. \cite{MarT99}). WFS
however, remains important despite the fact that it is a weaker logic
than that of stable models.  WFS provides a semantics for {\em any}
program including those with function symbols or those that have no stable
model.  WFS also has a low-polynomial data complexity~\cite{VG89}.
This lower complexity means that (1) programs do not need to be
grounded to be evaluated.\footnote{Section~\ref{sec-nongr-naf}
discusses how \Ergo addresses non-ground negative literals during an
evaluation.}; and (2) when tabled evaluation (tabling) is used, a
query $Q$ can be evaluated in a top-down manner using only relevant
clauses.


%
Before describing WFS, consider first the behavior of SLDNF
(cf. \cite{Lloy84}), the resolution method used by Prolog.  If
negation is not used, SLDNF resolution is simply Horn Clause
resolution with a fixed literal selection strategy, which in Prolog is
always left to right.\footnote{A {\em literal selection strategy}
refers to the order of resolution for literals in the body of a rule.}
SLDNF also supports a type of {\em negation as failure} via the Prolog
operator \verb|\+| (called \naf, in \Ergo syntax).  However, SLDNF
will not terminate for the query ``{\tt ?- s.}'' to any of the
programs in Figures~\ref{fig:table-examps1} or
\ref{fig:table-examps2}, all of which contain positive and/or negative
loops.

In contrast to SLDNF, \Ergo fundamentally relies on \SLGIA resolution,
which evaluates WFS via tabled {\em SLG
  resolution}~\cite{CheW96,CSW95} extended to provide {\em R}eactivity
to updates, and -- optionally -- {\em A}bstractions for bounded
rationality.  In this section we focus on SLG, which \Ergo uses to
support WFS as well as to adjoin argumentation theories to a program
(Section~\ref{sec:defeas}).  Reactivity of tabling is discussed in
Section~\ref{sec:dynamic}, while tabling with bounded rationality is
discussed in Section~\ref{sec:brat}.

In fact, \Ergo automatically uses tabled evaluation for queries to any
predicate not marked as transactional.\footnote{Since SLDNF is the
default evaluation strategy in Prologs that support tabling, a tabling
declaration is required to ensure termination and correctness for the
query to {\bf P$_{reach}$}.}  The formal operational semantics of SLG has
been presented in numerous papers (cf. \cite{SwiW12} for an overview);
in this paper we present tabling informally, in order to maintain
focus on \Ergo.  SLG resolution includes three mechanisms necessary to
evaluate queries according to WFS: {\em loop detection}, {\em
  incremental completion} for sets of mutually dependent subgoals, and
an {\em adaptive literal selection strategy}.\footnote{To fully
support WFS, SLG also performs other operations such as simplification
and answer completion that are not discussed here.}

We illustrate loop detection through evaluation of the query {\tt
  reachable(1,?Node)} to the definite Datalog program {\bf P$_{reach}$} in
Figure~\ref{fig:table-examps1}.  {\bf P$_{reach}$} is in \Ergo's {\em
  predicate syntax}, a subset of its \hilog syntax, as explained in
Section~\ref{sec:hilog}.
\begin{figure}[hbt] 
  \begin{tt}
    \begin{tabbing}
      fo\=fooo\=ooooooooooo\=ooooooooooooooo\=foo\=f\=fooooo\=\kill
  \hspace{0cm}\rule{0.75\textwidth}{0.5pt} \\
  ${\bf P_{reach}:}$ \\
\>  reachable(?X,?Y):- reachable(?X,?Z),edge(?Z,?Y) \\
\> reachable(?X,?Y):- edge(?X,?Y) \\
\> edge(1,2). \> \> edge(2,3).\>  edge(3,1).\\
  \hspace{0cm}\rule{0.75\textwidth}{0.5pt} 
    \end{tabbing}
  \end{tt}
  \caption{\Ergo Program to Illustrate Positive Recursion Handling in
    Tabling}\label{fig:table-examps1}>
  \end{figure}

  \noindent
This syntax is similar to that of Prolog except that variables in
\Ergo are denoted by identifiers that begin with {\tt ?} rather than
with an upper case letter.

Whenever SLG evaluates a subgoal $S$, it checks whether a variant of
$S$ is already present as a table entry.\footnote{Two terms are
variant if they are equal up to variable renaming. \Ergo uses a type
of tabling called {\em call variance} but also has an experimental
version that uses {\em call subsumption}, which can be faster for some
goals.} If not, as when $S$ is the top-level goal,
a table entry for $S$ is created, program clauses are used for
resolution, and any derived answers will be added to the table entry
for $S$.  Otherwise, if a table entry already exists, as when the goal
{\tt reachable(1,?Z)} is encountered as a body literal, the evaluation
sets up a {\em consumer subgoal} for $S$, and uses the tabled answers to
resolve against $S$. By guaranteeing that all answers for a subgoal
$S$ are returned to all consumer subgoals for $S$, tabling's loop
detection is sufficient to ensure correctness for queries to programs
like {\bf P$_{reach}$} that have no negation, but do have the {\em
  bounded term size} property in which all subgoals and answers have
a bounded size.

\begin{figure}[hbt] 
  \begin{tt}
    \begin{tabbing}
      fo\=fooo\=ooooooooooo\=ooooooooooooooo\=foo\=f\=fooooo\=\kill
\hspace{0cm}\rule{0.65\textwidth}{0.5pt} \\
  ${\bf P_{unfounded_{1}}:}$ \\
  \> s:- \naf p,\naf q,\naf r.      \>\>\> p:- q,\naf r.            \\
  \> q:- r,\naf p.        \> \>\>  r:- p,\naf q.      \\      
  \hspace{0cm}\rule{0.65\textwidth}{0.5pt}  \\
  ${\bf P_{undefined}:}$ \\
  \> s:- \naf p,\naf q,\naf r.\>\>\> p:- \naf r.            \\
  \> q:- \naf p.        \> \>\>  r:- \naf q.      \\      
  \hspace{0cm}\rule{0.65\textwidth}{0.5pt}  \\
  ${\bf P_{unfounded_{2}}:}$ \\
  \> s:- \naf p,\naf q,\naf r.\>\>\> p:- \naf r,q.            \\
  \> q:- \naf p,r.        \> \>\>  r:- \naf q,p.      \\      
  \hspace{0cm}\rule{0.75\textwidth}{0.5pt} 
    \end{tabbing}
  \end{tt}
  \caption{\Ergo Programs to Illustrate WFS and SLG}\label{fig:table-examps2}
  \end{figure}

For programs that contain negation such as {\bf P$_{unfounded_1}$} in
Figure~\ref{fig:table-examps2}, goal-directed evaluation of a query
such as {\tt s} also requires {\em incremental completion}.  To
support incremental completion, XSB's tabling engine implicitly
maintains a dependency graph, $G_{dep}$, among {\em active} subgoals.
$G_{dep}$ is a signed, directed graph in which there is an edge from
subgoal $S_1$ to subgoal $S_2$: if $S_1$ (directly) {\em depends on}
$S_2$: i.e., if there exists a substitution $\theta$ for a rule $R$ in a
program $P$ where $S_1$ is a variant of the head of $R\theta$ and
$S_2$ is a variant of the underlying subgoal of a literal $L$ in the
body of $R\theta$.  The edge is signed as positive or negative
depending on whether $L$ is a positive or negative literal.  A {\em
  strongly connected component (SCC)} in $G_{dep}$ indicates a set of
mutually dependent subgoals.  In the evaluation of {\bf
  P$_{unfounded_1}$}, the set of mutually dependent subgoals, {\tt
  p,q,r} forms an SCC.  A tabling engine with incremental completion
detects when all tabling operations have been performed for all
subgoals in an SCC as well as for all dependent subgoals.  At
completion, the status of the subgoals in the SCC ({\tt p,q,r}) is
changed from {\em active} to {\em complete} and the SCC is removed
from $G_{dep}$.  Since no subgoal in this SCC has any answers, and
since all dependencies within the SCC are positive, the subgoals form
an {\em unfounded set}.\footnote{In WFS, infinite dependency chains
also form unfounded sets. \Ergo's approach to such unfounded sets will
be discussed in Section~\ref{sec:brat}.} Their truth values are set to
\false, after which the subgoals {\tt \naf p}, {\tt \naf q} and {\tt
  \naf r} can all be resolved leading to the success of {\tt s}. Since
      {\bf P$_{unfounded_1}$} has a 2-valued well-founded model and
      since all unfounded sets can be discovered using a fixed
      ordering of a literal selection strategy, {\bf
        P$_{unfounded_1}$} is {\em fixed order dynamically stratified}
      \cite{SaSW99}.

The next two programs in Figure~\ref{fig:table-examps2}, {\bf
  P$_{undefined}$} and {\bf P$_{unfounded_2}$}, demonstrate the
importance of SLG's flexible literal selection strategy.  {\bf
  P$_{undefined}$} is the same as {\bf $P_{unfounded_1}$} except that
all positive literals have been removed from rule bodies.  In this
case, evaluation of the goal {\tt s} again discovers that {\tt p},
{\tt q} and {\tt r} form an SCC with no answers derived for any of the
goals.  However, because there are negative dependencies among these
subgoals, they do not form an unfounded set as before.  In such a
case, SLG {\em delays} evaluation of literals involving the negative
cycle ({\tt \naf p, \naf q} and {\tt \naf r}).  Delayed literals are
conceptually represented using a bar (\texttt{|}) in a partially
evaluated rule body.  Literals to the right of the bar have been
delayed, while literals to the left of the bar have not yet been
evaluated.  The state of a tabled evaluation can be partially
represented by the trees for each active subgoal.  Since {\tt p},{\tt
  q}, and {\tt r} are all active, and since each active subgoal has a
single rule, the state of the evaluation after these delays can be
schematically represented as:
\begin{tabbing}
fooooooooooooooooooo\=fooooooooooooooooooo\=oooooooooooooooooooo\=fooooo\=\kill
{\tt p:- \naf q|}. \> {\tt q:- \naf r|}. \> {\tt r:- \naf p|}.
\end{tabbing}
At this state, there are no other applicable operations for any of the
three active subgoals, so SLG adds the answers below to the tables of the
subgoals.  In the operational semantics of SLG, answers like these
that have a non-empty set of delayed literals correspond to atoms that
are undefined in WFS with truth value {\em u}.

The third program, {\bf P$_{unfounded_2}$} changes the order of
positive and negative body literals from {\bf P$_{unfounded_1}$}.  In
this case, {\tt p}, {\tt q} and {\tt r} will again form an unfounded
set, but each occurs to the right of a negative literal.  As in {\bf
  P$_{undefined}$}, SLG handles this by delaying negative literals for
the goals in the SCC.  The state after these delays is represented by
the partially evaluated clauses:
\begin{tabbing}
fooooooooooooooooooo\=fooooooooooooooooooo\=oooooooooooooooooooo\=fooooo\=\kill
{\tt p:- \naf q|r}. \> {\tt q:- \naf r|p}. \> {\tt r:- \naf p|q}.
\end{tabbing}
\noindent
Once the literals have been delayed, the evaluation continues,
discovering the unfounded set.  At this stage no operations are
available for these subgoals, and they are completed.  Since they are
completed with no answers, SLG marks them as \false, leading to the
success of the goal {\tt s} as in {\bf P$_{unfounded_1}$}.

Our purpose in this section is simply to review some aspects of SLG
that are used later in the paper.  Several complete descriptions of
SLG can be found in the literature (see the references in the discussion of
tabling in \cite{SwiW12}).  While the programs in
Figure~\ref{fig:table-examps2} might seem contrived, they represent
the types of evaluations \Ergo must perform for its argumentation
theories (Section~\ref{sec:defeas}).  Finally, we note that \Ergo also
uses extensions of SLG to evaluate transaction logic statements
(Section~\ref{sec:dynamic}) and to ensure sound termination of queries
to programs with function symbols via bounded rationality
(Section~\ref{sec:brat}).


       \section{Frame Logic} \label{sec:frames}

Frame Logic (abbreviated F-logic)~\cite{KiLW95} offers a number of benefits
for representing knowledge using a style that is both object-oriented
and ontology-oriented.  Although \Ergo can make use of external
knowledge represented in nearly any format, a pure \Ergo program
consists of rules and facts formed from a mixture of HiLog predicates and
F-logic {\em frame formulas} or {\em frames}.  In
this section we highlight frames along with {\em classification
  predicates} to denote a hierarchy of classes and instances, and then
describe how the two together provide structural and behavioral
inheritance for a program.  We should note here that the syntax used
in \Ergo differs from the original F-logic \cite{KiLW95} in many
respects. That original syntax was first improved by the \Flora
system~\cite{flora2-odbase-2003,flora2site} and then further extended
in \Ergo.

Throughout this paper, \Ergo features are illustrated using a set of
examples motivated by the US DARPA AIDA project.\footnote{{\tt
  www.darpa.mil/program/active-interpretation-of-disparate-alternatives}}
In 2021-2022 the AIDA project constructed sample problems centered
around tracking misinformation about the COVID-19
virus~\cite{DARPA-AIDA}.  To address these problems, various natural
language inputs were analyzed and the results represented as a set of
knowledge graphs
over which various types of reasoning and
hypothesis generation were performed.\footnote{Names and
organizations in the examples are motivated by the comic strip
``Washingtoon'' popular in the 1980s and 1990s.}

\subsection{Frames and Inheritance}

Figure~\ref{fig:instances} shows two \Ergo frames, each with
an {\em object-id}, such as \texttt{person('Bunky Muntner')} or {\tt
  claim(13355)}, followed by a list of attribute-value
pairs of the form $\langle attr \rangle$ \texttt{->} $\langle val
\rangle$. For instance, the types of the {\tt claim} attribute values
include atoms, an RDF XSD date type instance, another frame expression
representing the semantic frame of a statement, and even the reference
to a Python object.  Figure~\ref{fig:instances} also shows a fragment
of the classification hierarchy for the program, where the infix
operator \texttt{\mysub/2} represents the subclass relation and
\texttt{\rawisa/2} indicates that an instance is a member of a
class.  Since their object-ids identify instances, these frames are called
{\em instance frames}.
\begin{figure}[hbt]
  \begin{tt}
    \begin{tabbing}
      foooo\=f00\=foofoofoofoooooofoofo\=ooooooooofpoofoo\=\kill
      {\color{purple}1} person('Bunky Muntner')[age->63, \\
        \> residesIn->zip(20016)[city->Washington,state->DC], \\
        \> occupation->journalist,\\
        \> spouse->'Ginger Muntner', \\
           {\color{purple}5}           \> language->Spanish].\\
      \> \\
~~claim(13355)[dateTime->\verb|"2021-12-23T12:33:55"^^\dt|,\\
          \>  sourceDocument$\mbox{->}$ 'Jalapeno Springs Daily', \\
          \>  medium -> NewsSite, \\
{\color{purple}10}        \>  sourceText -> 'My vitamin supplement cures everything.'\\
          \>  statement -> statement1[disease->COVID,cure->takeSupplements ], \\
          \>  embedding -> pyObj(p0x7faca3428510)]. \\
        \\
~~       Person \mysub Agent. \>\> \>~~~~ 
Organization \mysub Agent. \\
{\color{purple}14}       person(?) \isa Person. \>\>\>~~~~ claim(?) \isa Claim. \\
~~        claim(13355) \isa NewsReport. 
      \end{tabbing}
    \end{tt}
\caption{\Ergo instance frames with a fragment of a classification hierarchy} 
\label{fig:instances}
\end{figure}
Both \texttt{\mysub/2} and \texttt{\rawisa/2} can be specified either
by facts or rules. For instance, the non-ground fact on line 14
indicates that any frame with an object-id that unifies with {\tt
  claim(?)} belongs to the class {\tt Claim}.\footnote{Throughout
this paper we adopt the convention that identifiers for classes begin
with an upper-case letter.}

Figure~\ref{fig:frametypes} shows \Ergo class frames that specify {\em
  signatures} for subclasses and instances, default values for
instance attributes, as well as attribute values of the class itself.
When an attribute value is a type, the attribute-value relation is
designated via \texttt{=>/2} and the inheritance is termed {\em
  structural}; otherwise \texttt{->/2} is used to designate default
object values, and the inheritance is termed {\em behavioral}.  Both
type and object values are inheritable if they occur within
$[|....|]$.  but otherwise are non-inheritable.  For instance, in the
\texttt{Claim} frame in Figure~\ref{fig:frametypes}, all attribute
values are inherited except for \texttt{author}, which pertains to
the \texttt{Claim} class itself, and has the object {\tt Ingrid
  B. Baird} as its value.  Also in the frame for \texttt{Claim} the
attribute \texttt{dateTime} is specified to be required and functional
through a {\em cardinality constraint}, which indicates the minimum
and maximum number of values an instance may have for a given
attribute.  The \texttt{Claim} frame also denotes the default object
value \texttt{X} for the attribute \texttt{medium}.  Finally, an \Ergo rule on
line 12 propagates to person instances values of {\tt claimed} from
{\tt claimer} attributes of claims.

\begin{figure}
    \begin{tt}
      \begin{tabbing}
      fooo\=foooooo\=f0ffoofoofoooooofoofooooooooooooof\=fooopo\=\kill
1\>        Claim[|dateTime\{1..1\}=>\dateTime,\   \>\>~~ Person[|age=>\integer, \\
\>\>          sourceDocument=>Document,    \>\>   spouse=>Person, \\
\>\>          documentType=>PublicWebDocument   \>\>   residesIn=>Location,\\
\> \>          sourceText=>AtomicConstant,  \>\>   politAffil=>PolitParty,\\
5\>\>           statement=> CompoundTerm,    \>\>   occupation=>Occupation,\\ 
\>\>           claimer=>Agent,              \>\>   claimed=>Claim, \\
 \>\>          embedding => PythonObject,   \>\>   language->English,\\
\>\>           medium->X|].                  \>\>  politAffil->Independent|].\\
\>        Claim[author->'Ingrid B. Baird']. \\
10\\
\>       NewsReport[|documentType->WebNewsArticle|]. \\
\>       ?Agt[claimed->?Claim]:- ?Agt:Agent,?Claim[claimer->?Agt].\\
\>       WebNewsArticle::PublicWebDocument.
      \end{tabbing}
    \end{tt}
  \caption{\Ergo class signatures} 
  \label{fig:frametypes}
\end{figure}

The instances in Figure~\ref{fig:instances} inherit these attributes in
various ways:
\begin{itemize}
\item {\em Direct behavioral inheritance.} Because the instance
  \texttt{'Bunky Muntner'} is a \texttt{Person}, it directly inherits
  \texttt{politAffil->Independent}.
\item {\em Behavioral overriding.} The claim object
  \texttt{claim(13355)} states explicitly that {\tt
    medium->NewsSite} (in Figure~\ref{fig:instances}),
  overriding the default {\tt medium->X}
  inherited from the class \texttt{Claim}.
\item {\em Monotonic structural inheritance.} \texttt{claim(13355)} is a member of both the class {\tt Claim}, from which it
  inherits the default {\tt documentType} as
  \texttt{PublicWebDocument}, and of class
  \texttt{NewsReport} where \texttt{documentType} has the type \texttt{WebNewsArticle}.
  Unlike behavioral inheritance of attribute values, structural type
  inheritance is not overridable.  Instead, it works
  \emph{monotonically}, by accumulation.  In our example, the class
  \texttt{WebNewsArticle} is said to be a subclass of {\tt
    PublicWebDocument} in Figure \ref{fig:frametypes},
  so the two types would be considered consistent for nearly all
  applications.  In general, if \Ergo derives different types for an
  attribute of an object, the types are not taken to be contradictory.
  This viewpoint is similar to that of formal ontologies in which two
  concepts are not taken to be disjoint unless a relevant exclusion
  axiom causes them to be.  In \Ergo, similar constraints can be
  enforced either by type checking or by integrity constraints (see
  Section~\ref{sec:flexsearch}).  The paper
  \cite{kifer-yang-inheritance-2006} gives rigorous treatment to the
  semantics of inheritance in \Ergo.
  \item {\em Monotonic behavioral inheritance.}  While behavioral
    inheritance is defeasible by default in \Ergo, there are cases
    when monotonic inheritance is preferred.  In our example the
    people described by
    the knowledge base are assumed to speak English, but they
    may speak other languages as well.  \Ergo allows behavioral
    inheritance to be set as either monotonic or non-monotonic 
    at the module level.  Figures \ref{fig:instances} and
    \ref{fig:frametypes} and the attribute \texttt{language} 
    illustrate this point. 
    The default strategy for behavioral inheritance is non-monotonic and
    thus the language (or languages) stated at Bunky
    Muntner's object level in Figure~\ref{fig:instances} overrides the
    inheritance of \texttt{language}  from class \texttt{Person} specified in
    Figure~\ref{fig:frametypes}.
    If, however, the module were told to use the monotonic semantics for
    behavioral inheritance (say, by inserting the directive \texttt{:-
    setsemantics\{inheritance=monotonic\}} then \texttt{language->English}
  would be inherited \emph{monotonically} from class \texttt{Person}, i.e. it
  would be added to the statement \texttt{language->Spanish}
  specified directly inside the object for Bunky Muntner.
Thus a query such as
\begin{alltt}
      ?- person('Bunky Muntner')[language->?Language].
\end{alltt}
would unify {\tt ?Language} with both {\tt English} and {\tt Spanish},
as desired.
\end{itemize}




\subsection{Attributes Defined via Rules} \label{sec:python}

Attributes can be defined by \Ergo rules, facts, or both. As an
example we show how \Ergo can represent and maintain contemporary
knowledge sources such as statement embeddings used by Language Models
(LMs).
%
Both LLMs and small LMs such as Phi-3~\cite{phi3} and Google's
  Universal Sentence Encoder~\cite{CYKH18}
  are usually written in Python and C$^{++}$ and so are easily
  accessible by \Ergo through the Janus interface~\cite{SwiA23}.  For
  instance, suppose embeddings were obtained using a version of
  Universal Sentence Encoder, whose Python API loads a model, and then
  calls it with different textual inputs.  The model object is created
  by the goal.
\begin{tabbing}
{\tt Py\_func(tensorflow\_hub,load($<model\_url>$),?PyObj)}
\end{tabbing}
\noindent
The predicate {\tt Py\_func/3} allows \Ergo to call a Python function,
in this case
\begin{tabbing}
  {\tt tensorflow\_hub:load($<model\_url>$)}.
\end{tabbing}

The return of this call unifies the \Ergo variable \texttt{?PyObj} with a  Python object reference of the form
{\tt pyObj(ObjectReference)}.  After
asserting the object reference as a {\tt currentModel/1} fact, a rule
for {\tt Claim} objects can be written as {\tt
    \begin{tabbing}
  fooo\=foof\=foofooooooooooooo\=foofooooooooooooo\=foo\=foo\=\kill
  \> ?Obj:claim[embedding->?Vector]:- \\
  \> \> currentModel(?Model), \\
  \> \> ?Obj[sourceText->?Text], \\
  \> \> Py\_dot(Model,embed(Text),?Vector).
\end{tabbing}
}
\noindent
where {\tt ?Vector} will be instantiated to a list of floating point
numbers.  Unlike this explanatory example, in practice it is
usually best to keep information such as embedding vectors as
attributes of Python objects, which can be accessed and manipulated  through Janus in
much the same manner as \Ergo objects.

\subsection{Type Checking}
A type in \Ergo can be any class in the classification
hierarchy. Remembering that the class membership (the \texttt{:/2}
relationship), the class hierarchy (the \texttt{::/2} relationship),
and the properties (the \mvd/2 relationships) can be defined via logic
rules, it is no wonder that \Ergo types are powerful and expressive
but also that fully static type-checking is not possible.
\emph{Dynamic} type checking \emph{is} possible.  While dynamic type
checking cannot be guaranteed to terminate and may be slow in
complex cases, it is often a highly useful aid to program
development.  To perform dynamic type checks, one can easily express
different type \emph{violation} as queries that return type
violations as \emph{counterexamples} to type correctness.

Although the semantics needed for signatures is sometimes
program-specific, many type inconsistencies
are common, including (1) that an object has an attribute value
inconsistent with the inferred type for that value; (2) an object has
an attribute that is not in any of its inherited signatures; or (3) the
number of values for a given object and attribute are not consistent
with cardinality constraints declared in an inherited signature.  Type
checking can be particularly important when complex objects are
inserted in a dynamic setting, as will be discussed in
Section~\ref{sec:dynamic}.

Here is a type violation query that identifies attributes and values
for instances that violate properties 1 and 2 above.
  \begin{alltt}
type_error(?Obj,?Attr,?R,?D) :-
       %% Values that violate typing
       ?Obj[?Attr->?R], ?Obj[?Attr=>?D], \naf ?R:?D
       ;
       %% Defined methods that do not have type information
      ?Obj[?Attr->?R], \naf ?Obj[?Attr=>?_D].

?- type_error(?Object,?Attr,?Result,?Class).
\end{alltt}

\noindent
To save the user the trouble of writing type constraints for common
cases of type violations, \Ergo provides the \typecheck module that
supports the above type checkers so that the user can simply query:
  \begin{alltt}
   Type[check(person('Bunky Muntner')[?Meth->?],?Result)]@\typecheck.
  \end{alltt}
Details can be found in the \Ergo user's manual.



       \subsection{\Ergo Frame Compilation.} \label{sec:frame-comp}
The compilation of \Ergo frames into Prolog predicates follows the
general design of \cite{KiLW95} and consists of two parts: (1) frames
are translated into Prolog; and (2) {\em closure rules} are
added to support the object-oriented semantics of frames.  In this
section, we describe a few aspects of \Ergo compilation to give the
reader an idea of the compilation process.

Both instance and class frames are translated into Prolog in a similar
manner, so we take as an example the following instance frame, which
is to be added to \Ergo's knowledge base.  The frame is {\em complex}
because it has multiple attribute-value pairs, because its identifier
includes an inheritance predicate (\isa/2), and because it has a
set-valued attribute.
\begin{alltt}
Mary \rawisa employee[age->29, salary(1998)->100000, kids->\{Tim,Leo\}]. 
\end{alltt}
\noindent
As described in \cite{KiLW95}, frame compilation begins with
translating {\em complex} frames (like the above) into a conjunction
of \emph{atomic} frames.  As part of this process, attribute values
that are sets are split into a conjunction of singletons, and the
membership statement for {\tt Mary} is factored out of the frame:
\begin{alltt}
Mary \rawisa employee,Mary[age->29],Mary[salary(1998)->100000],
                   Mary[kids->Tim],Mary[kids->Leo].
\end{alltt}
\noindent
These atomic frames are then translated to Prolog facts that are
asserted to the XSB database as {\em tries}~\cite{RRSSW98}.
\begin{alltt}
isa(Mary,employee).
mvd(Mary,age,29).                   mvd(Mary,salary(1998),100000).
mvd(Mary,kids,Tim).                 mvd(Mary,kids,Leo).
\end{alltt}
The {\em wrapper predicate} {\tt mvd/3} (which stands for multi-valued
data) is used to translate attributes with behavioral
inheritance.  Query frames must undergo the same translation as
program frames for resolution to work properly.  For instance the
query
\begin{alltt}
  Mary:?Class[kids->?Kid].
\end{alltt}
would be translated to
\begin{alltt}
isa(Mary,?Class),mvd(Mary,kids,?Kid).
\end{alltt}
yielding two solutions, one with {\tt
  \{?Class=employee,?Kid=Leo\}} and another with {\tt
  \{?Class=employee,?Kid=Tim\}}. \Ergo's compiler mangles the names of the
wrapper predicates to include the module of the file or query.

\Ergo rule frames  must be encoded in a more complex manner than fact frames,
and require generation of {\em closure rules} to:
\begin{itemize}
\item Compute the transitive closure of \mysub /\texttt{2} in a manner that
  supports runtime warnings about cycles in the subclass hierarchy.
\item Compute the closure of \isa\texttt{/2} with respect to \mysub
  \texttt{/2}, i.e., \\ if {\tt X \isa C} and {\tt C \mysub D} then {\tt X \isa
    D}.
\item Perform inheritance. In particular, monotonic inheritance must
  detect and propagate multiple type values while non-monotonic
  inheritance must detect and override object value conflicts.

\item Provide an infrastructure to treat undefined methods as
 exceptions.  Unlike Prolog, undefined predicates in \Ergo do not
  throw errors by default.  However when checks for undefinedness are turned on, code
  to raise appropriate errors is injected into the closure rules.
\end{itemize}

The above closure rules are specified statically in special files that
get included in the compiled Prolog code after the appropriate
preprocessing that makes the files compliant with \Ergo's dynamic
module system.


       \section{HiLog} \label{sec:hilog}

The previous section showed how complex frames, called {\em frame
  formulas}, can be constructed and used, but the examples also
contained logical terms such as the object id {\tt claim(13344)} and
the occurrence of {\tt py\_func/3} in the body of a rule.  \Ergo fully
supports HiLog~\cite{CKW93} which gives a higher-order syntax to
Prolog-style resolution.  The idea behind \hilog can be explained
easily.
For instance, the \hilog fact 
 \begin{tabbing}
{\tt p(fact(7))(a,f(b,b),[3,?X])}
  \end{tabbing}
\noindent
is schematically translated to Prolog as:
\begin{tabbing}
{\tt apply(apply(p,fact,7),a,apply(f,b,b,[3,?X]))}.
\end{tabbing}
where every non-list compound term $T$ is translated into a term with
functor {\tt apply/n} and $T$ as its first argument.  By similarly
translating a query such as
\begin{tabbing}
{\tt ?- p(?X)(a,f(b,b),[3,foo])}
\end{tabbing}
to
\begin{tabbing}
\texttt{?- apply(apply(p,?X),a,apply(f,b,b),[3,foo])},
\end{tabbing}
%
the query can be resolved against the fact with either tabled or
non-tabled Prolog-style resolution.  To be more specific a \hilog term
is defined as follows and this semantics is equivalent to the
model-theoretic view of HiLog \cite{CKW93}.

\begin{definition}
  A \emph{\hilog atomic constant} is an atom.  A \emph{base \hilog term}
  is a \hilog atomic constant, a list, a variable, or an arithmetic
  expression.  


  A \emph{general \hilog term} is a base \hilog term, or has the form
$T(\overline{Args})$ where $T$ is a general \hilog term, and
  $\overline{Args}$ is an argument list of general \hilog terms.
\end{definition}

\hilog is sometimes said to have a higher-order syntax but a
first-order semantics.  Specifically \hilog semantics is based on a
translation into Prolog.

\begin{definition} \label{def:hilog-transform}
  The \emph{\hilog transform} $tr_H$ is defined as follows.

  If $Term$ is a base \hilog term, then $tr_H(Term) = Term$.  Otherwise
  if $Term = Term_1(\overline{Args})$ then

  \[tr_H(Term) = apply(tr_H(Term_1),tr_H(\overline{Args}))\]
\noindent
  where $tr_H(\overline{Args})$ simply applies $tr_H$ to each
  argument in the sequence.
\end{definition}
\noindent
It can easily be seen that by applying $tr_H$ to a \hilog term
produces a Prolog term.  As a result, if $tr_H$ is applied both to
\hilog programs and goals in a program $\cP$, the model of $\cP$
coincides with model of its translation. Operationally, we refer
informally to resolution of translated goals as {\em \hilog
  resolution}.

Since variables can appear nearly anywhere in HiLog terms,
HiLog is useful for exploring the structure of knowledge
bases.
Another use case is traditional higher-order programming.
Suppose we need to transitively close many different
binary relations, like \texttt{flight/2}, \texttt{edge/2},
\texttt{parent/2}. Some of these needs may arise in the future and not be
known at present. Instead of writing the usual 2-rule   definition of the
transitive closure in each such case,
we could write just one pair of rules for all current and
future needs:
\begin{verbatim}
   closure(?P)(?F,?T) :- ?P(?F,?T).
   closure(?P)(?F,?T) :- closure(?P)(?F,?Mid),?P(?Mid,?T), 
\end{verbatim}
Note the implicit reliance on tabling for the second, left-recursive
rule.  Now, to compute the transitive closure of \texttt{flight/2} and
of \texttt{edge/2} we can simply pose these two queries:
\begin{verbatim}
   ?- closure(flight)(?F,?T).
   ?- closure(edge)(?F,?T).
\end{verbatim}

\noindent
\textbf{Avoiding name clashes:} In systems that are implemented via a
translation to a target system, name clashes with other components of
the target system become a concern. This consideration applies not only to
the HiLog \emph{wrapper predicate} \texttt{apply} but also to frame-wrappers
\texttt{mvd/3}, \texttt{isa/2},
etc. (cf. Section~\ref{sec:frame-comp}).  Such unintended interactions
are usually hard to detect and debug.  \Ergo addresses this problem by
\emph{salting} (also called mangling) the wrapper predicates with strings of
characters that are unlikely to occur under normal
circumstances. Thus, instead of \texttt{apply}, \texttt{mvd}, etc.,
the salted versions of these wrappers are used: \saltapply, \saltmvd, \saltisa, etc.,
where \emph{salt} denotes the string  that includes rare characters such as the
quotes, \$, and various others. For instance, \Ergo
uses \verb|'_$_$_ergo'| to salt the wrapper predicates used in
HiLog/F-logic-to-Prolog translations.

\section{The Module System of  \Ergo } \label{sec:modules}
In \hilog and F-logic, the number of unique predicates used for the
Prolog translations is small.  As a result, it is easy to write a goal that
backtracks through all clauses in the program, a behavior that can
lead to
inefficiency.  The salting
technique of the previous section prevents unintended resolutions
between \Ergo's wrapper predicates and similarly named predicates from
other XSB programs that may be used in conjunction with \Ergo in the same
XSB process. Without salting there could be name
clashes between the \texttt{mvd} and \texttt{apply} wrappers discussed
in previous sections and \texttt{mvd} or \texttt{apply} used in other
XSB programs (if XSB predicates are in the default \texttt{usermod} module).
But there is another problem: the separation of namespaces among the
wrappers \emph{within} the \Ergo system.  Such separation is necessary
to enable component-wise construction of large knowledge bases and to
reduce the chances of inadvertent interactions.
As in most languages, \Ergo's
module system addresses these and similar issues.

\Ergo's module system was designed to be dynamic and to
support a number of other unique features. Modules can be created ``on
the fly'' at run time, and rules and data can be loaded into them,
inserted, or deleted at run time as well. Calls from one module to
another may also be constructed at run time.  At the same time, any
module can prevent unwanted calls by encapsulating itself.
Encapsulation can happen at compile time or dynamically, via the
executable
\texttt{export/1} directive (e.g., \texttt{export\{?[abc \mvd ?],
  pqr(?)\}} or even \texttt{export\{?[abc \mvd ?], pqr(?)\} >>
  module1,module2}).
Furthermore, modules need not be instantiated within a goal; for instance 
one can query the modules and argument values for which \texttt{p/1} is
true:
\begin{verbatim}
    ?- p(?X)@?M.
\end{verbatim}
Some of the above features exist in various logic programming systems, but
others (variables over modules, support for non-predicate-based languages
like HiLog and F-logic) are unique to \Ergo.

The implementation of \Ergo modules elaborates on the idea of salting
the wrappers such as \texttt{apply/n} and \texttt{mvd/3}.  The goal is
to ensure that wrappers used in \emph{different} modules are
translated using \emph{different} salted wrappers.  For example, the
statements \texttt{John[age->11]} and \texttt{Mary[age->12]} that
appear in the \emph{same} module would be related to each other
because they would be encoded using the same set of salted wrappers:
like \texttt{\saltmvd}, \texttt{\saltisa}, or \saltapply.  In
contrast, these same statements in \emph{different} modules would be
encoded using \emph{different} salted wrappers.

\subsection{Module-aware Salting}
Thus, the basic idea underlying the implementation of \Ergo modules is
that in the end, all \hilog and F-logic wrapper predicates are translated
with respect to a particular module, and the module symbol is used
only in the outermost functor of the translated wrapper predicates.
This restriction is made because ultimately the system determines the
truth values of the instances of the wrapper predicates only.  For instance,
the module-specific version of HiLog's {\tt apply/n} is
\texttt{\saltapplymod/n+1}.
Functors at inner levels become arguments to the outer wrappers---
after the \hilog and F-logic transforms, inner functors are no longer
treated as predicates, have no truth values, and their salting does
not involve modules. As a result, the arguments conceptually play the
role of shared object identifiers about which logical assertions
(facts and rules) are being made. Assertions made in one module about
a set of objects do not interact with the assertions about that same
set of objects in another module unless one module explicitly queries
the other.

For example: suppose a database table named \texttt{p/1}  exists both  in module
\texttt{foo} and module \texttt{bar}. A priori we do not know of any
relationship between these two tables so their contents should be
considered to be unrelated. However, if we learn that module \texttt{foo}
has a rule of the form   
\begin{tabbing}
~~~~
\texttt{p(?X) :- p(?X)@bar.}   
\end{tabbing}
then the version of \texttt{p/1} in \texttt{foo} must necessarily have
all the tuples that the version of   
\texttt{p/1} in \texttt{bar} has. 

Admittedly, salting only the outer functor of a term means 
that we  distinguish between \texttt{p(?X)} that appears as a call
and \texttt{p(?X)} that appears as an argument,  which makes meta-programming 
noticeably less elegant. This trade-off is well-known and analogous to
choosing between a Prolog with
a predicate-based module system and a Prolog with a functor-based module
system (e.g., Ciao Prolog vs. XSB).

Because assertions are module-specific, queries are issued from a
particular module, although they may involve querying 
another module. For instance,
\begin{tabbing}
~~~~
{\tt ?- p(?X)(a,f(b),[3,?X]), q(?X)@bar.}
\end{tabbing}
translates into a query of the form

\begin{tabbing}
  \texttt{?-
    '\textit{salt}\^{}\textit{modname}\^{}apply'('\textit{salt}\^{}apply'(p,\_X),a,'\textit{salt}\^{}apply'{}(f,b),[3,\_X]),}
  \\
  ~~~~~~\texttt{'\textit{salt}\^{}bar\^{}apply'(q,\_X).}
\end{tabbing} 
\noindent
Similarly, the wrapper predicates \saltmvdmod, \saltisamod, etc. are used to separate translations 
of the frame statements into different modules.
For rules, the translation is similar:
\begin{tabbing}
  ~~~~
  \texttt{hd(?X) :-  ?Y:?X@bar, foo::?Y.}\\
  \textnormal{becomes}\\
  ~~~~
  \texttt{\textit{salt}\^{}\textit{modname}\^{}apply'(hd,\_X) :-}
  \\
  ~~~~~~~~~~~
  \texttt{'\textit{salt}\^{}\texttt{bar}\^{}isa'(\_Y,\_X),}
  \texttt{'\textit{salt}\^{}\textit{modname}\^{}sub'(foo,\_Y).}
\end{tabbing}
Note that after the compilation of the above query,
the module-part of the first outermost wrapper predicate,
\texttt{\textit{salt}\^{}\textit{modname}\^{}apply'/4},
remains generic, i.e., it is not replaced with a real
module name. Similarly, in the above rule,
the module parts of the wrappers 
\texttt{\textit{salt}\^{}\textit{modname}\^{}apply'/2} and
\texttt{'\textit{salt}\^{}\textit{modname}\^{}sub'/2}
remain generic.
As may be guessed, this happens because
the above goal \texttt{p(?X)(....)}  does not have any module
reference (@\emph{mod}); likewise, neither \texttt{hd(...)} nor
\texttt{foo::?Y} in the above rule has such references.  So, what is
the module in those cases and how would the compiler obtain that
module's name?  In \Ergo, the answer is that the module in question is
the ``\emph{current}'' module, a concept closely tied to the dynamic
manner in which rules and modules are created and associated with each
other during the course of a computation.

\subsection{Loading vs. Adding}
Normally the user creates a file that includes facts, rules, and
queries.  Then the user loads or adds that file to a module (the
module may or may not exist at the time), which causes the file to be
compiled \emph{for that module} and the queries executed. The user
might pose additional queries, load or add additional files into
new or existing modules, or perform other operations.

For concreteness, suppose we have a file called
\texttt{myProgram.ergo} that contains the query and the rule 
just discussed plus some data facts (i.e., bodiless rules).  We can
load the file into some explicit module, e.g., \texttt{mainRules}, or into the
\emph{current} module:
\begin{verbatim}
  ?- [myProgram >> mainRules].
  ?- [myProgram].
\end{verbatim}
The first form will create the empty module \texttt{mainRules}, if it
did not exist before, compile \texttt{myProgram.ergo} for the module
\texttt{mainRules} causing compiled salted Prolog rules to appear
in XSB's memory.  (If the module existed prior to the command it is
\emph{emptied out} of its old contents first.)
In
the second version of the load command,
\texttt{myProgram.ergo} will be compiled for the \emph{current} module
but the rest is the same as we just explained.

The current module dynamically changes during an \Ergo session.  When \Ergo
starts up, it automatically creates the module \texttt{main} which is the first current module.
If either of the above load commands executes next, it
executes with \texttt{main} as the current module.  During the
compilation of \texttt{myProgram.ergo} using the first option, the
current module will be \texttt{mainRules} rather than {\tt main}.
With the second version of the load command, the current module will
not change, of course, and will still be \texttt{main}.  If
\texttt{myProgram.ergo} includes another loading command, like
\texttt{?- [myStuff >> miscRules]}, then when this command is
executed, \texttt{myStuff.ergo} will be compiled with
\texttt{miscRules} as the current module.
   
Loading wipes out any existing contents of the target module, so
loading two different sets of rules into the same module is rarely
meaningful unless the first set of rules and data has become obsolete.
Such scenarios arise when dealing with streaming data, but are
unlikely in cases where working knowledge of an application evolves. For
such applications, \Ergo provides the \emph{add} command, e.g.,
\begin{verbatim}
  ?- [+myAddlStuff >> mainRules].
  ?- [+myAddlStuff].
\end{verbatim}
Using the add command resembles loading except that it does not erase
beforehand the contents of the module (\texttt{mainRules} in the first
form, the current module if the second form).  The overall result of \Ergo's
dynamic creation of modules, loading, and adding is that any given
file can either be loaded into any module or added at any
time.\footnote{In Prolog, files may be conditionally loaded into a
module, but this approach is somewhat more restrictive than in \Ergo.}




\subsection{Compiling Dynamic Module References}
Until now, we remained silent about how the
placeholder for the current module, \emph{modname}, gets replaced with
the actual name of that module within the context of a given
execution.
This
is accomplished with the help of the GPP
preprocessor \footnote{\texttt{https://github.com/logological/gpp}}
a core program used by XSB as well. GPP is similar to the CPP
preprocessor used by C and C++, but includes features needed by
\Ergo that CPP does not support such as the ability to change syntax
on the fly and to perform macro evaluations inside quoted strings.

Because code files may be dynamically loaded or added into modules,
\Ergo does not compile into the XSB Prolog syntax directly
but rather into a syntax where HiLog and frame wrappers are enveloped
with suitable GPP macros.  For example, instead of compiling the
\texttt{apply/n} wrapper into
\texttt{'\textit{salt}\^{}bar\^{}apply'(q,\_X)} and
\\ \texttt{'\textit{salt}\^{}\textit{modname}\^{}apply'('\textit{salt}\^{}apply'(p,\_X),a,'\textit{salt}\^{}apply'{}(f,b),[3,\_X])}
directly, we could use these GPP macros:
\begin{verbatim}
#mode save
#mode nostring "\!#'"
   #define  PREFIX                      _$_$_ergo
   #define  SALT_WRAP(BaseWrap)         'PREFIX^BaseWrap'
   #define  MODSALT_WRAP(Mod,BaseWrap)  'PREFIX^Mod^BaseWrap'
   #define  CURRMOD_WRAP(BaseWrap)      'PREFIX^CRNT_MODULE^BaseWrap'
#mode restore
\end{verbatim}
and compile our HiLog terms into these more structured and compact
expressions:
\begin{verbatim}
MODSALT_WRAP(bar,apply)(SALT_WRAP(apply)(q,_X)) 
CURRMOD_WRAP(apply)(SALT_WRAP(apply)(p,_X),a,SALT_WRAP(apply)(f,b),[3,_X])
\end{verbatim}
There are several interesting things to note about these macros.
First, the last three macro definitions refer to other macros such as
\texttt{\#PREFIX}, and these referenced macros appear inside quoted
strings.\footnote{As far as we know, the CPP preprocessor in C/C++ cannot be
forced to evaluate such macros, but XSB's GPP can.} The \texttt{\#mode
  save/restore}
commands above order GPP to temporarily stop interpreting
single quotes as string delimiters, which enables GPP to compute all
those macros as we intended. Second, notice that \texttt{CRNT\_MODULE}
is undefined as a macro, but it is clear from the name that we somehow
want it to evaluate to the current module name (which recall, 
changes depending on the context where the reference to the current
module appears). In our example, the current module can be
\texttt{main} or \texttt{mainRules} and this becomes known only at the
time of loading.  Fortunately, GPP can be forked from its parent
process and can take a directive to add a new macro
\begin{alltt}
  #define  CRNT\_MODULE   \emph{module-being-loaded} 
\end{alltt}
where \texttt{\emph{module-being-loaded}} is the module name known to the
loader. In other words, the problem of passing the name of the current
module  is solved  because we can pass that name to GPP under the guise of
the \texttt{CRNT\_MODULE} macro.

As a final point, \Ergo's module system also extends so that Prolog
modules are called in a manner similar to \Ergo modules.
For instance, the \Ergo goal that invokes
the Prolog predicate \texttt{member/2}, which lives in the XSB module
{\tt basics}, is:
\begin{verbatim}
   member(?X,?List)@\prolog(basics).
\end{verbatim}
This is implemented simply by compiling \texttt{member/2} without any
salting whatsoever. Since the core of XSB is written in C, C routines
appear as regular Prolog calls, so \Ergo can access them in
the same way, via the \verb|@\prolog| idiom.  
In principle, this idiom could be extended also to
other languages so that, for example, Python modules would be
accessible via \verb|@\python| or \verb|@\janus|, if Janus is used.


       \section{Rules and Defeasibility} \label{sec:defeas}
As mentioned in Section~\ref{sec:prelim},
\Ergo's default negation (\naf) is based on WFS.
In Section~\ref{sec:frames} we also described how \Ergo supports both
defeasible behavioral inheritance and monotonic inheritance of
signatures (i.e., types).  At this point we turn our attention to \Ergo's
support for explicit negation and several versions of defeasible
reasoning in the LPDA style \cite{Grosof97,WGKFL09}, each defined by
its own \emph{argumentation theory.}
In fact, any properly defined argumentation theory for \Ergo can be
enabled in a user module, and different such theories can be active at
the same time as long as they are in different modules.

An argumentation theory in \Ergo acts behind the scenes, providing the
ability to extend the base SLGR(A) resolution strategy of the system
in a well-defined, declarative way.  \Ergo's default argumentation
theory is {\em Generalized Courteous Logic Programming (GCLP)}
introduced in \cite{Grosof97,WGKFL09}, which \Ergo further extends
with \emph{cancellation} along with other refinements developed in the
more than fifteen years since those publications.  A typical
argumentation theory in \Ergo is visible to a user through three
predicates: \opposes{}, \overrides{}, and \cancel{}. It is through
this interface that a user can control how defeasibility is handled by
the reasoning system.
We first describe the overall effect of defeasibility on \Ergo's
inference through an example, next we examine the above predicates for
defeasibility control, and then we discuss their semantics.

\begin{figure}[hbt]
  \begin{minipage}{8.5in}
\begin{alltt}
:- use\_argumentation\_theory.  \textcolor{blue}{// Use the default argumentation theory}
\textcolor{blue}{// Defeasible rules with rule descriptors}
\textcolor{Emerald}{@!\{pr1[author->Swift,isa->PotencyRule]\}}
\textcolor{Green}{@\{Potency(High,?Per,?V)\}}
?Per[believes->hasPotency(?V,High)] :-
        ?V:Virus,
        ?Per[residesIn->?Loc],
        ?\_Claim:Claim[
           claimer->?Per,
           statement->\$\{?Loc[shouldRequire->SchoolClosure[purpose->reduceXmit(?V)]]\}].
\textcolor{Emerald}{@!\{dr1[author->Kifer,isa->DefaultPotencyRule]\}}
\textcolor{Green}{@\{Potency(Low,?Per,?V)\}}
?Per[believes->hasPotency(?V,Low)]:- ?Per:Person, ?V:Virus.
\textcolor{blue}{// Defeasibility control statements}
\bs{}opposes(Potency(?L1,?Per,?V),?\_AnyGoal1, Potency(?L2,?Per,?V),?\_AnyGoal2) :-
        ?L1 != ?L2, ?V:Virus, ?Per:Person.
\bs{}overrides(Potency(High,?\_Per,?\_V), Potency(Low,?\_Per,?\_V)).
\bs{}cancel(Potency(?,?,?)):- \bs{}neg Pandemic.  \textcolor{blue}{// Cancel Potency rules if no pandemic}
\textcolor{blue}{// Some other (strict) rules}
?Per[residesIn->?GPE]:- ?Per[residesIn->?Loc], ?Loc[city->?GPE].
?Per[residesIn->?GPE]:- ?Per[residesIn->?Loc], ?Loc[state->?GPE].
\textcolor{blue}{// Base Facts}
ECM:Virus.
person('Bunky Muntner'):Person
 [ age->63,
   residesIn->zip(20016)[city->Washington,state->DC],
   spouse->'Ginger Muntner'].
claim(12345):Claim
 [ medium-> Instagram,
   topic->'Government Containment Actions: School Closures',
   location->'Washington DC',
   statement->\$\{zip(20016)[shouldRequire->SchoolClosure[purpose->reduceXmit(ECM)]]\},
   claimer->person('Bunky Muntner')].
\textcolor{blue}{// The query: what do the people believe should be done?}
\textcolor{magenta}{?-}  ?Per:Person[believes->hasPotency(?V,?Strength)].
\end{alltt}
    \end{minipage}
    \caption{A Fragment of a Belief Logic} \label{fig:belief}
\end{figure}

  Figure \ref{fig:belief} shows a fragment of a set of beliefs
  about claims that we assume are loaded into the
  module {\tt Belief}.
  The directive \texttt{:- use\_argumentation\_theory} at the top ensures that 
  \Ergo's default argumentation theory
  will be activated when this fragment of the knowledge
  base is loaded into a module.
  The LPDA support enhances the syntax of \Ergo with the ability to
  optionally attach
  \emph{rule descriptors}   to each rule.
   In our example they are colored in shades of green and appear as
  \textcolor{Emerald}{{\tt @!\{...\}}} and \textcolor{Green}{\texttt{@\{...\}}} constructs.
  The \textcolor{Emerald}{\texttt{@!\{...\}}} portion of the descriptor assigns a
  unique \emph{rule id} and provides meta-data about the rule; if no
  rule id is explicitly associated with a rule by the author, one is
  created automatically.
  %
  The \textcolor{Green}{\texttt{@\{...\}}} portion of the rule
  descriptor is a {\em rule tag}, which makes the rule {\em defeasible}
  (as opposed to {\em strict}, i.e., non-defeasible). These tags are
  used heavily in defeasible reasoning.\footnote{Rule tags can
  alternately be written by placing the attribute-value pair
  \texttt{tag->Potency(...)} but in many cases the short form
  \texttt{@\{\textnormal{\textit{SomeTag}}\}} of the descriptor is
  more convenient than the general form
  \texttt{@!\{\textnormal{\textit{SomeId}}[...,tag->\textnormal{\textit{SomeTag}},...]\}}}

  Rule descriptors provide meta-data for rules by means of frames that
  can be used much in the same way as other \Ergo frames. In
  Figure~\ref{fig:belief}, the rule descriptors have attributes such
  as \texttt{author} and {\tt comment} that resemble those of other
  rule systems.  However, by querying rule descriptor frames using the
  construct {\tt @!\{...\}}, a rule can be linked into a
  classification hierarchy in which, for example,
    \begin{tt}
      \begin{tabbing}
        fooooooooooooooooooooo\=foofooooooooooooo\=foofooooooooooooo=\kill
        defaultPotencyRule::PotencyRule.  \>\> PotencyRule::ViralRule.
      \end{tabbing}
    \end{tt}
One easy way to do this is by adding the following rule to the {\tt Belief}
module:
  \begin{tabbing}  
    \texttt{?RuleId\isa ?Class :- @!\{?RuleId[isa->?Class]\}.}
  \end{tabbing}
  
  Rule tag descriptors (\texttt{\textcolor{Green}{\texttt{@\{...\}}}}) are used
  by \Ergo's argumentation theories.  In general, a tag descriptor can
  contain any valid frame or \hilog term.  The rule with the tag {\tt
    Potency(High,?Per,?V)} says that a person believes the potency of
  a virus is \texttt{High} if the person has made a public claim that
  their local schools should be closed. The rule with the tag {\tt
    Potency(Low,?Per,?V)} serves as a default rule.

The syntax \$\{...\} -- seen for instance in the body of the low
potency rule -- \emph{reifies} logical formulas such as frames
into objects that can be used as values of attributes or as arguments
of HiLog terms and predicates.  In this example, reification allows 
the representation of the beliefs of different voters about school closing.

The three \emph{defeasibility control predicates} in the middle of
Figure~\ref{fig:belief} extend athe GCLP argumentation theory to be
domain-specific.  \opposes/2 provides conditions for (resolvable)
contradiction
beyond the implicit contradiction that always exists between an atom
{\em A} and its explicit negation \eneg A.  In this example the
opposition statement enforces the semantic constraint that a rational
person should not believe that the same virus has different overall
potencies.  In addition to marking rules as defeasible, the rule tags
of Figure~\ref{fig:belief} are also used for prioritization via
\overrides.
In our case, the \overrides/2 fact ensures
that, for a given person and virus, any other potency rule overrides
the \texttt{Low} potency rule.  Note that by using the tags,
argumentation predicates can be coded to apply to precise sets of
rules or derived heads. Finally, the \cancel rule ensures that all
potency rules are disregarded by the belief logic unless there is an
exogenously defined state of a pandemic.

At this point, it is instructive to try to understand at an intuitive
level what defeasibility of the first two rules in
Figure~\ref{fig:belief} has given us.  If we ask the query shown as
the last statement in that figure:
\begin{verbatim}
     ?- ?Per:Person[believes->hasPotency(?V,?Strength)].
\end{verbatim}
we get:
\begin{verbatim}
   ?Per = person('Bunky Muntner')  ?V = ECM   ?Strength = High
\end{verbatim}
which seems an answer we intuitively expect. However, if we focus on
just the rules and the base facts, while ignoring the defeasibility
control predicates, then our knowledge base appears to imply two beliefs:
\begin{verbatim}
   person('Bunky Muntner')[believes-> hasPotency(ECM,High)]
   person('Bunky Muntner')[believes-> hasPotency(ECM,Low)]
\end{verbatim}
where the first fact is derived via the first rule in the figure (with
Id \texttt{pr1}) and the second fact by the second rule with Id
\texttt{dr1}.  On the other hand, since \texttt{person('Bunky Muntner')
  : Person} and \texttt{ECM : Virus},
\begin{tt}
\begin{tabbing}
  fooooooooo\=ooooooooooo\=foofooooooooooooo\=foofooooooooooooo=\kill
  \opposes(Potency(High,person('Bunky Muntner'),ECM),\\
        \> Potency(Low,person('Bunky Muntner'),ECM))
\end{tabbing}
\end{tt}
is also derived, so that the above two belief statements cannot both
be true.  The \texttt{\bs{}overrides} statement says that the
derivations made via rule \texttt{pr1} (high potency) are more
important than derivations made by rule \texttt{dr1} (low potency).
As a result, the answer of the belief in low potency above gets
discarded and our knowledge base returns only the first (high-potency)
answer.

As an experiment, one could try to comment out the defeasibility control
predicates and pose the same query again.
Now we get \emph{two} answers, consistent with what the 
normal derivation in \Ergo yields (sans defeasibility).
\begin{verbatim}
   ?Per = person('Bunky Muntner')  ?V = ECM   ?Strength = High
   ?Per = person('Bunky Muntner')  ?V = ECM   ?Strength = Low
\end{verbatim}
So, clearly defeasibility control predicates have an effect.  This
effect is the result of \Ergo augmenting the knowledge base with an
\emph{argumentation theory} that specifies what to do if one kind of
rule overrides the other, one kind of derived fact conflicts with
another, etc. During normal operation, the argumentation theory is
applied behind the scenes and is completely transparent to the
user.\footnote{On the other hand, the actions of opposition,
overriding and cancellation can all be examined when debugging.}

\subsection{The GCLP Argumentation Theory} \label{sec:argumentation}
Pseudo-code for \emph{GCLP}, \Ergo's default argumentation theory is
shown in Figure~\ref{fig:PseudoArg}.  The actual code in \Ergo is more
complex due to optimizations and debugging support.\footnote{GCLP in
\Ergo also allows strict rules to override defeasible rules in cases
of opposition.}  That code also exports a number of query points that
let the user investigate why certain rules were defeated, refuted, or
rebutted; why there is a conflict between some rules; etc.
%
As mentioned, \Ergo supports several argumentation theories other than
GCLP that differ in how rules are defeated, and in the
generality of opposition.  For instance, some argumentation theories
allow \opposes/2 rules to specify when a set of three or more
different rules must succeed for opposition to happen.

Regardless of the argumentation theory, a defeasible rule:
\begin{tabbing}
\emph{@\{RuleTag\} Head  :- Body}
\end{tabbing}
is translated to 
\begin{tabbing}
\emph{@\{RuleTag\} Head :- Body, \naf defeated(RuleTag,Head)@AT.}
\end{tabbing}
where \emph{AT} is the \Ergo module that contains the desired
argumentation theory. Observe that this simple cross-module calling
scheme enables different \Ergo modules to use different argumentation
theories at the same time.  A defeasible rule with variable bindings
for which {\em Body} succeeds is called a {\em candidate}

\begin{figure}[tbhp] 
  \begin{tt}
  \begin{footnotesize}
\begin{tabbing}
  fooo\=of\=foooooofoooooooooooo\=foofooooooooooooo\=foo\=foo\=\kill
defeats(?Tg1,?Hd1.?Tg1,?Tg2):- refutes(?Tg2,?Hd2,?Tg1,?Hd1). \\
defeats(?Tg1,?Hd1.?Tg1,?Tg2):- rebuts(?Tg2,?Hd2,?Tg1,?Hd1). \\
\> \\
defeated(?Tg1,?Hd1):- refutes(?Tg2,?Hd2,?Tg1,?Hd1). \\
defeated(?Tg1,?Hd1):- rebuts(?Tg2,?Hd2,?Tg1,?Hd1). \\
defeated(?Tg1,?Hd1):- disqualified(?Tg1,?Hd1).\\
\>\\
rebutted(?Tg2,?Hd2):- rebuts(?,?,?Tg2,?Hd2). \\
\> \\
rebuts(?Tg1,?Hd1,?Tg2,?Hd2) :- \\
\> \> conflicts(?Tg1,?Hd1,?Tg2,?Hd1), \\
\> \> \naf refuted(?Tg1,?Hd1). \\
\> \\
refuted(?Tg2,?Hd2):- refutes(?,?,?Tg2,?Hd2). \\
\> \\
refutes(?T1,?Hd1,?Tg2,?Hd2) :- \\
\>\>  conflicts(?Tg1,?Hd1,?Tg2,?Hd2),\\
\> \> \overrides(?Tg1,?Hd1,?Tg2,?Hd2), \\
\> \> \naf refuted(?Tg1,?Hd1). \\
\> \\
conflicts(?Tg1,?Hd1,?Tg2,?Hd2) :- \\
\>\>  \opposes{}(?Tg1,?Hd1,?Tg2,?Hd1), \\
\>\>  candidate(?Tg1,?Hd1),candidate(?Tg2,?Hd2). \\
\>\>  \naf\ \cancel{}(?Tg1,?Hd1), \naf\ \cancel{}(?Tg2,?Hd2). \\
\> \\
candidate(?Tg,?Hd) :- getBody(?Tg,?Hd,?Body), call(?Body). \\
\> \\
disqualified(?Tg2,?Hd2):- transitively\_defeats(?Tg1.Hd1,?Tg2,?Hd2),\\
\> \> \> \naf rebutted(?Tg1,?Hd1).\\
disqualified(?Tg,?Hd):- \cancel{}(?Tg,?Hd).         
\end{tabbing}
  \end{footnotesize}
  \end{tt}
  \caption{Simplified Pseudo-Code for GCLP, \Ergo's Default Argumentation Theory}
  \label{fig:PseudoArg}
\end{figure}
 It can be seen from Figure~\ref{fig:PseudoArg}
that a given rule, say $R_{t}$ with tag {\em Tag} (denoted
$R_{t}/Tag$) is defeated if it is refuted or rebutted by some
non-defeated rule, or if $R_{t}/Tag$ is disqualified.  Rebuttal
requires that an opposing rule, say $R_{o}$, with tag {\em OppTag}
(denoted $R_{o}/OppTag$) succeeds as a candidate, and that
$R_{o}/OppTag$ is neither canceled nor refuted.  Refutation on the
other hand also requires an appropriate \overrides/2 statement to be
present.  Thus in GCLP if two undefeated rules rebut each other but neither
is overridden, both fail.  Otherwise the \texttt{overrides} statement
causes either one rule to succeed and the other to fail, or both to
succeed with truth value $u$ in the case of mutual overriding.
Rule sets like that in Figure~\ref{fig:PseudoArg}, when optimized for
performance, implement \Ergo's argumentation theories.

We should note that the literature is rife with flavors of
defeasible logic
\cite{Brewka00prioritizingdefault,Delgrande03aframework,dung-argument-01,eiter-preference-2003,GelfondS97,Nute94,Prakken93-annals,SakamaI00,wang-lin-00,ZhangWB01},
each slightly different from the next.
It is argued in \cite{WGKFL09,wan-lpda-asp14} that argumentation theories,
like those in \Ergo, can simulate and approximate a number of other
approaches, and the dozen or so  argumentation theories supplied with \Ergo
testifies to the power and flexibility of that approach.


       \subsection{Argumentation and Paraconsistency in \Ergo}
The following example highlights a property of \Ergo's GCLP
argumentation theory: the failure of rebutted rules.
\begin{example}[\em Explicit Negation in GCLP]
  \label{ex-eneg}
Given the preceding discussion, it is useful to consider the semantics tbat GCLP
assigns to the opposition of defeasible atoms {\tt A} and {\tt \eneg A} in
Program $P_{eneg}$ 
  \begin{tt}
    \begin{tabbing}
      fo\=fooo\=ooooooooooo\=ooooooooooooooo\=foo\=f\=fooooo\=\kill
      {\normalsize${\bf P_{eneg}:}$} \\\vspace{1mm}
\> ~~:- use\_argumentation\_theory \\
\> \\
\> ~~@\{tag1\} A. \> \> @\{tag2\} \eneg A. 
    \end{tabbing}
  \end{tt}
Here $A$ and \eneg $A$ are both facts and thus candidates.  Since they
have tags, they are defeasible.  GCLP and many other LPDA-style
argumentation theories make an implicit assumption that \texttt{A} and
\texttt{\eneg A} always oppose each other. Due to the use of GCLP
in Program $P_{eneg}$ the defeasibility transformation ensures that
every defeasible rule $R$ is made to guard against being defeated
(i.e., the literal \emph{\naf defeated(RuleTag$_R$,Head$_R$)@AT} is
added to the body of $R$). Therefore, by the \texttt{refutes/4} rule in
Figure \ref{fig:PseudoArg}, they conflict. By the \texttt{rebuts/4}
rule in Figure \ref{fig:PseudoArg} we see that {\tt A} and
\texttt{\eneg A} rebut each other. Since neither of these facts is
refuted, they are both defeated, so that both {\tt A} and
\texttt{\eneg A} are false.
\end{example}

GCLP's treatment of explicit negation is the same as most of
\Ergo's argumentation theories, but it differs from approaches
like WFSX~\cite{AlDP94} in which both {\tt A} and {\tt \eneg A} would
have the truth value {\em undefined}.  The fact that GCLP sets their
truth value to \false makes it arguably more useful for \emph{paraconsistent}
reasoning based on frame logic with argumentation.  If the truth value
of opposing literals were set to {\em undefined} then the truth values
of literals that depend on these opposing literals would also be set
to {\em undefined}.

Although some logics use a four-valued lattice to distinguish unknown
from incompatible truth values, like those in $P_{eneg}$, handling
inconsistency purely within WFS requires mapping inconsistency to an
existing truth value: \true, \false, or \undefined.  Viewed in
this way, WFSX conflates the inconsistency of a literal with a
literal's being undefined/unknown (\undefined) while GCLP conflates it with \false.
In fact, \Ergo users have suggested a way to support a four-valued
truth lattice semantics based on an argumentation theory that acts
like GCLP
{\em except} that it has no concept of rebuttal. This argumentation
theory, \texttt{RefuteCLP}, is included with \Ergo. In
\texttt{RefuteCLP}, if one rule refutes another, any inferences drawn
by the second rule are defeated. However, if two rules rebut (but do
not override) each other then neither inference is defeated and it is
possible to infer that, in a pair of opposing facts, both are true.
Therefore, if $P_{eneg}$ in Example \ref{ex-eneg} used RefuteCLP instead of
GCLP, both \texttt{A} and \texttt{\eneg A} would have been true.   
\Ergo does not provide
direct support for the fourth truth value, \emph{inconsistent}, but
user-built rulebases can treat such situations as \emph{localized
inconsistency}. Localized inconsistency is similar to paraconsistency
in Annotated Logic \cite{BlSu89,KiLo92}, and sophisticated users
can employ meta-interpreters or similar techniques to detect 
and take advantage of paraconsistency in \Ergo knowledge bases.

       a\section{Flexible Search Strategies for Unsafe Subgoals} \label{sec:flexsearch}
As mentioned in Section~\ref{sec:prelim} \Ergo does not require a
program to be ground to be evaluated, a major factor in its
scalability.  However, this freedom can give rise to {\em unsafe}
queries that do not satisfy a required instantiation pattern.  Given
\Ergo's reliance on well-founded negation within non-monotonic
inheritance and argumentation, unsafe negation must be addressed.  In
addition, \Ergo's use of external knowledge may also give rise to
unsafe queries that must be addressed.

\Ergo's approach to unsafe queries uses {\em instantiation delay}
together with the WFS {\em u} truth value.  Instantiation delay is
different from the \SLGIA delay presented in Section~\ref{sec:prelim},
although the two mechanisms do bear some resemblance. As discussed,
\SLGRA delay is invoked on a ground negative literal $G_T$ when the
tabling completion operation encounters a loop through negation
involving $G_T$.  Afterwards, $G_T$ is never re-evaluated, but instead
simplified away when further evaluation of goals upon which $G_T$
depends determines it to be \true or \false.  On the other hand,
instantiation delay occurs when a goal $G_I$ is not instantiated
enough to be evaluated; afterwards, any time a variable in $G_I$ is
instantiated, $G_I$ is rechecked to determine whether it satisfies an
instantiation property (such as nonvar or ground), and evaluated if
so.

In this section, we first show how \Ergo automatically uses
instantiation delay to address unsafe negation, and then how it can be
invoked by the user to handle safety requirements imposed by external
knowledge sources, or for other reasons such as improving query
performance.

\subsection{Unsafe Negation in \Ergo} \label{sec-nongr-naf}
Consider the resolution of the goal {\tt p(?X)} using the set of rules
and facts in Figure \ref{fig:neg-nonground}. Since by default
\Ergo uses a left-to-right literal selection strategy, the literal selected
in Step 1 of the resolution is \texttt{q(?X)}, which is then resolved
using a program rule to produce the subgoal \texttt{\naf q2(?X)}.
Since \texttt{\naf q2(?X)} is an unsafe literal, \Ergo attempts to
\emph{delay} the literal's evaluation in the hope that
\texttt{?X} will be grounded later on in the evaluation---for example,
the grounding of \texttt{?X} might happen after resolving
\texttt{r(?X)} later in the rule for {\tt p(?X)}.

The postponement is done using a modified version of the {\tt when/2}
predicate found in XSB and other Prologs, which in this case delays
the goal {\tt \naf q2(?X)} until {\tt ?X} becomes ground.  This is
shown on line 3 of the evaluation.  The predicate \texttt{when/2} is
based on {\em attributed variables}, which underlie the evaluation of
logical constraints in Prolog systems.
\begin{figure}[hbtp]
  \longline
    \begin{tt}
      \begin{tabbing}    
        fooo\=fooo\=fooofooo\=fooofoooofoooo\=fooofoooofoooo\=foo\=foo\=\kill
 {\bf P$_{unsafe}$} \\ \vspace{1mm}\\
\>        p(?X):- q(?X), \naf s(?Y), r(?X),
 t(?Z,?Y).\\
\>        q(?X) :- \naf q2(?X).\\
\>        r(?X) :- r2(?X).\\
\> q2(1).  \>\>      r2(2).\>       s(3).  \>     t(4,?). \\
  \hspace{0cm}\rule{0.75\textwidth}{0.5pt}  \\
 \textbf{Evaluation of \textnormal{\tt p(?X)}:} \\
{\color{purple}1}\>        p(?X):- q(?X), \naf s(?Y), r(?X), t(?Z,?Y).\\
\>        p(?X):- \naf s(?Y), r(?X), t(?Z,?Y), \>\>\>\>{\color{blue}// \textnormal{delay \tt q2/1}}\\
{\color{purple}3} \>      \> ~~~~~when(ground(?X),\naf q2(?X)). \\
{\color{purple}4} \>        p(?X):-  r(?X), t(?Z,?Y),  \>\>\>\>{\color{blue}// \textnormal{also delay \tt s/1}}\\
\>      \> ~~~~~when(ground(?X),\naf q2(?X)), when(ground(?Y),\naf s(?Y)). \\
        {\color{purple}6} \>        p(?X):-  r2(?X), t(?Z,?Y), \>\>\>\> {\color{blue}//
          \textnormal{ \texttt{r2/1}  binds
  \texttt{?X}  to 2}} \\
\>      \> ~~~~~when(ground(?X),\naf q2(?X)), when(ground(?Y),\naf s(?Y)). \\
{\color{purple}8} \>        p(2):-  \naf q2(2), t(?Z,?Y), when(ground(?Y),\naf s(?Y)). \\
\>        p(2):-  t(?Z,?Y), when(ground(?Y),\naf s(?Y)). \\
{\color{purple}10} \>        p(2):-  not\_exists(s(?Y)). {\color{blue}//
  \textnormal{fallback to XSB's well-founded \texttt{not\_exists/1}}} \\
\>        No  \> \>\>\> {\color{blue}// \textnormal{fail, since \texttt{s(3)}  is \true}}
      \end{tabbing}    
    \end{tt}
  \longline
  \caption{Schematic Evaluation of Non-Ground Default Negation}
 \label{fig:neg-nonground}
\end{figure}
In the next step, \texttt{\naf s(?Y)} is selected and postponed, as
shown on lines 4--5.  Next, \texttt{r/1} is resolved with the third
rule and then with \texttt{r2(2)}, which finally binds \texttt{?X} to
2, grounding the negative subgoal \texttt{\naf q2(?X)} to {\tt \naf
  q2(2)}. This is shown on lines 6--8. Since {\tt \naf q2(2)} is \true,
we remove it, reaching line 9.  Note that grounding of \texttt{\naf
  q2(?X)} was requested on line 3 but did not happen until line 8,
after \texttt{r2(?X)} was resolved. The two terms involved in this
interaction, \texttt{\naf q2(?X)} and \texttt{r2(?X)}, do not even
occur in the same rule. Moreover, when \texttt{\naf q2(?X)} was
delayed, it was not known whether \texttt{?X} would ever become ground.
The point is that the algorithm for subgoal delay is automatic, and
involves inter-rule interaction and run-time analysis.

Resolving \texttt{t(?Z,?Y)} leaves 
a non-ground subgoal \texttt{\naf s(?Y)}, which cannot be grounded
later because the variable \texttt{?Y} does not appear outside of that
subgoal.  How should one proceed? In our experience, in most such
cases, the user wants to check whether \emph{no grounding} of
\texttt{?Y} would ever make \texttt{s(?Y)} \true.
This can be efficiently verified using the XSB predicate
\texttt{not\_exists/1}, as shown on lines 10 and 11 in
Figure~\ref{fig:neg-nonground}.
Roughly, XSB defines
\texttt{not\_exists(G)} as \texttt{(TG, fail ; tnot(TG))}, where {\tt
  TG} is a tabled version of \texttt{G}, and \texttt{tnot/1} is XSB's
ground well-founded negation.
In certain cases, the user might want to assign the truth value \emph{u}
to non-ground \naf literals, such as \texttt{\naf s(?Y)}. This option
can be selected in \Ergo via a runtime switch.

Our discussion above considerably simplifies \Ergo's actions; for
instance the \naf operator also handles variables in \emph{quantified}
subgoals, based on extensions of \cite{LlT84}.  Furthermore, it is
generally accepted that \texttt{\eneg G} implies \texttt{\naf G}
\cite{AlDP94}, and \Ergo also uses this inference when grounding
\naf-negated subgoals.  We note that many aspects of \Ergo's approach
to \naf could be adopted by Prologs that implement WFS, such as SWI
and XSB itself.  In fact, nu-Prolog \cite{NaDZ89} implements subgoal
reordering for Prolog negation, although it does not assign the truth
value {\em undefined} to non-ground negative subgoals, nor does it
handle explicit negation or quantifiers.



\subsection{Delay Quantifiers} The delay quantifiers
{\tt wish/1} and {\tt must/1} cause dynamic reordering of subgoals
that do not fulfill argument binding requirements. These quantifiers
check conditions on variables in a subgoal and delay that subgoal if
these conditions fail.  For example, the subgoal
  \begin{alltt}
must(ground(?X) or nonvar(?Y))\^{}?X[foo->?Y]
    \end{alltt}
contains the delay quantifier {\tt must/1}, and evaluation of the
subgoal delays evaluation of {\tt ?X[foo->?Y]} until either {\tt ?X}
is ground or {\tt ?Y} is partially instantiated. If the desired
instantiation is unachievable, a runtime error results. The
\texttt{wish/1} quantifier delays its subgoal in the same manner as
\texttt{must/1}, but the subgoal is executed after no more delay is
possible, and no error is raised.


       \section{KRR in a Dynamic Setting} \label{sec:dynamic}
Most real-world knowledge changes over time.  A \texttt{Claim}  object, like that
in Figure~\ref{fig:instances}, may be added or modified.  New rules
expressing beliefs, like those in Figure~\ref{fig:belief}, may be
designed by a knowledge engineer or learned.  Knowledge must also
adapt to changing circumstances.  For instance, if polls show that a
viral pandemic is not a high concern of voters, a number of rules may
be canceled in the sense of defeasible logic
(Section~\ref{sec:defeas}).

In \Ergo, {\em every} rule and fact is treated as dynamic information,
so that any rule or fact may be inserted, deleted, enabled, or
disabled.\footnote{ Unlike in Prolog, where every rule/fact is either
dynamic (deletable) or static, based on predicate-level declarations.  }
The changes made possible by such flexibility can give rise to
different issues.  A new claim may need to be added
\emph{transactionally} so that either all necessary elements must be
added or none may be.  Adding or canceling a rule can have complex
consequences especially when argumentation theories are used or when a
change is made to a program's classification hierarchy.  As a result,
assurance may be needed that certain conditions hold or do not hold
after an update.
%
\Ergo addresses such issues through Transaction Logic
\cite{BoKi94}
extended with integrity constraints.\footnote{\Ergo also supports two other mechanisms for monitoring or managing change: latent queries, and alerts.}

Returning to our running example about claims and beliefs, it may be
useful to distinguish between two classes of people who frequently
make claims: politicians and journalists.  In fact, we might like to
ensure that no person is both a politician and a journalist within the
time frame covered by our knowledge base.  It is easy to write a
rule to check for this:
\begin{tabbing}
foof\=oofooooooooooooo\=foofooooooooooooo\=foo\=foo\=\kill
{\tt occupationConstr(?Per):- } \\
\>   {\tt ?Per:Person,?Per[occupation->journalist],?Per[occupation->politician].}
\end{tabbing}
How can this constraint be enforced?
%
One might try defeasible reasoning, as in
Section~\ref{sec:defeas}, and set the two occupations in opposition,
but that could result in such persons having no occupation whatsoever
(depending on the argumentation theory).
In such cases, it is usually better to impose database-style \emph{integrity
constraints} via this command:
\begin{tabbing}
foofoofooooooooooooo\=foofooooooooooooo\=foo\=foo\=\kill
{\tt ?- +constraint\{occupationConstr(?)\}.}
\end{tabbing}
This activates a database constraint that tells \Ergo to check the query
\begin{tabbing}
foofoofooooooooooooo\=foofooooooooooooo\=foo\=foo\=\kill
{\tt
  ?- occupationConstr(?Person).}
\end{tabbing}
after each transaction.  In logic programming, such a constraint is
considered to be violated if this query returns any answers: such
answers are treated as objects that \emph{violate} the constraint. In
our case by executing
\begin{tabbing}
foofoofooooooooooooo\=foofooooooooooooo\=foo\=foo\=\kill
\texttt{?- insert\{'Bunky Muntner'[occupation->politician]\}.}
\end{tabbing}
the above query returns a {\tt Person} object that has
\texttt{journalist} and \texttt{politician} both listed as occupations
thereby violating the constraint.
Various options are available to handle such cases --- from simple
warnings to transactional rollbacks to callbacks to routines for
handling the constraint violation.
%

Transactional rollbacks happen automatically, if updates are written using
\emph{Transaction Logic} \cite{BoKi94}.  
At the practical level, this means that the user must use
\emph{transactional}  update operators,
i.e., \verb|t_insert|, \verb|t_delete|,
\verb|t_enable|, \verb|t_disable|, etc.,
instead of \texttt{insert},  \texttt{delete}, \texttt{enable},
\texttt{disable}, etc.
For instance,
\begin{verbatim}
?- t_insert{'Bunky Muntner'[occupation->politician]}. 
\end{verbatim}
Operationally, one can think of such updates as \emph{backtrackable},
which provides the rollback functionality.  Transactional updates are
committed when a goal returns to the interpreter level, or when {\tt
  commit@\verb|\db|} is executed.  A transactional update may be part
of a complex query that relies on extensive tabling. To address this,
whenever a transactional update is executed or rolled back (i.e.,
backtracked over), any inferences that have been made are reactively
adjusted via \Ergo's \SLGRA, which uses incremental tabling
(cf. \cite{Swif14}), ensuring {\em view consistency} of inferences with
respect to updates.

As a second example, consider a toy  \Ergo program for graph coloring:
%
  \begin{tt}
    \begin{tabbing}
foof\=oofo\=oooooooooooooooofoooofoooooooooooooooo\=foo\=foo\=\kill
      ${\bf P_{Color}:}$ \\
\> colorGraph :- \naf uncoloredNode(?).   \> \>
              {\rm /* stop if all nodes are colored */} \\
   \> colorGraph :- colorNode, colorGraph.
                \> \> {\rm /* color graph recursively */} \\
\> colorNode :-                          \> \> {\rm /* color one node */} \\
\> \> \naf coloredNode(?N), \\
\> \> color(?C)    \\
\> \> \naf (adjacent(?N,?N2),colored(?N2,?C)), \\
\> \> t\_insert(colored(?N, ?C)). 
    \end{tabbing}
  \end{tt}
%
\noindent
{\bf P$\_{color}$} colors a graph by recursively picking a node $N$,
picking a color for $N$, and then ensuring that $N$ is not adjacent to
a node of the same color.  If the adjacency constraint cannot be
satisfied, the program backtracks to find a new color, or to recolor a
previous node.  In the latter case, updates to {\tt colored/2} will be
rolled back as necessary for the search to continue.  (Of course,
graph coloring is most efficiently done by a constraint system or
SAT-checker.)  Within \Ergo, backtracking through updates is more
expensive than normal backtracking, which is performed efficiently by
XSB's engine.  However, the paradigm of using backtrackable updates
together with incremental tabling can be useful for problems where the
search space is highly complex and not well understood, such as
analysis of malicious object code~\cite{SCDGHH18}.\footnote{This work
was begun before \Ergo's transaction logic was stable, so the paper
uses XSB's \SLGI resolution directly, with the application program
itself logging and rolling back updates.}

At the theoretical level, Transaction Logic is a higher-order logic with a
model and proof theory.
Specifically, it treats elementary updates, like insert
and delete, as oracles, so that the approach can be applied to
first-order logic with belief revision as well as to WFS with fact or
rule updates.  In either case, satisfiability is defined over {\em path
  structures} consisting of a sequence of program states separated by
update operations.  Thus semantically, finding a solution for a goal
to {\bf P$_{color}$} amounts to finding a path structure that is a
model for the program and goal.\footnote{In this and other respects,
Transaction Logic, which models data updates and belief
revisions, differs from modal logics such as dynamic
logic~\cite{Pratt78} that model accessibility of program states using
a semantics of Kripke structures.}  The result is that data updates
can be used in an \Ergo program like {\bf P$_{color}$} without
compromising \Ergo's logical semantics.

%

\Ergo's support for management of changing knowledge has a number of
other features such as bulk transactional updates and rule updates.
In fact, from the user's point of view, \Ergo handles transactional
updates of rules in nearly the same manner as facts, supporting
transactionality, view consistency, integrity constraints and other
features.  However there is one difference.  Because some \Ergo rules
are compiled into static Prolog code that cannot be updated, \Ergo
gets hold of the rules not by their syntax but rather through their
rule ids (cf. Section~\ref{sec:frames}). For this purpose, it provides two
predicates, \texttt{t\_enable\{\textnormal{\emph{RuleId}}\}} and
\texttt{t\_disable\{\textnormal{\emph{RuleId}}\}}, 


       \section{Termination and Bounded Rationality in \Ergo.} \label{sec:brat}
As discussed in Section~\ref{sec:prelim} \Ergo's use of \SLGI
resolution ensures termination and produces the correct well-founded
model for programs with the bounded term depth property.  This
termination class includes Datalog and is far larger than the
termination class of Prolog using SLDNF.  However, assuming queries are
side-effect free, runtime \Ergo options can guarantee correct
termination for any query to a program that has a finite model, and
{\em sound} termination on any query to any program.

\subsection{Termination on Programs with Finite Models}
Section~\ref{sec:prelim} provided examples of SCCs of mutually
dependent subgoals that correspond to unfounded sets in WFS.  Another
type of unfounded set is an {\em unfounded chain}, which can be
illustrated by the simple \Ergo program:
  \begin{tt}
    \begin{tabbing}
      foo\=foo\=f\=fooooo\=\kill
      ${\bf P_{FinModel}:}$ \\
      \> p(?X) :- p(f(?X))
    \end{tabbing}
  \end{tt}
\noindent
Using \Ergo's default resolution method, \SLGI, the goal {\tt p(a)}
to $P_{FinModel}$ produces an infinite chain of goals, \texttt{p(f(a))},
\texttt{p(f(f(a)))}, \texttt{p(f(f(f(a))))} and so on.  This is
because \SLGI creates a new table entry for any goal that is not a
variant of a goal already in the table (cf. Section~\ref{sec:prelim}).
\Ergo can address such situations via {\em subgoal abstraction}, in
which subgoals are abstracted once they reach a maximum size.  In
$P_{FinModel}$, abstraction with subgoal depth 2 or more would rewrite
any goal that unified with {\tt p(f(f(?X)))} to {\tt p(f(?X))},
ensuring that the evaluation will create only a finite number of
subgoals.

The use of subgoal abstraction guarantees powerful semantic
properties.  Let us represent a three-valued well-founded model $M$
in terms of two sets: the set of atoms that are \true in $M$ and the set of
atoms that have truth value \undefined in $M$.  Since the well-founded
model of $P_{FinModel}$ has no atoms with truth value \true or
\undefined, its three-valued model is finite.  In \cite{RigS14}, it is
shown that if SLG is extended with subgoal abstraction then it
correctly terminates for any program with a well-founded model that is
finite under the above definition.

In \Ergo, subgoal abstraction can be requested with the help of the
\texttt{goalsize} 
restraint in the \texttt{setruntime} directive. Loading into \Ergo  the small fragment
below, 
\begin{verbatim}
   ?- setruntime{goalsize(4,abstract)}.
   p(?X) :- p(f(?X)).
   p(b).
   p(f(c)).
   ?- p(?X). // the query
\end{verbatim}
produces the following answers and terminates:
\begin{verbatim}
   ?X = b
   ?X = c
   ?X = f(c)
\end{verbatim}
Without subgoal abstraction ({\it i.e.}, if the directive
\begin{tabbing}
      fofo\=f\=fooooo\=\kill
\>  \texttt{setruntime\{\linebreak[1]goalsize(4,abstract)\}}
\end{tabbing}
\noindent
is commented out) the above program will not terminate and will return
no answers.  In our case, however, the \texttt{goalsize} parameter
tells the query subsystem to abstract subgoals larger than size 4,
which forces termination.\footnote{The actual size is computed by
taking into account the function symbols in the goal with their
arities in a prefix-style traversal that preserves termination
properties.}  Subgoal abstraction is implemented by XSB's tabling
subsystem.  {\revision Whenever a tabled goal is encountered, a 'subgoal
check/insert' operation traverses the goal in order to determine
whether an existing table may be used or a new table must be created.
The declaration for subgoal abstraction activates an engine-level hook
that abstracts the goal as part of the traversal~\cite{RigS14}.}

\subsection{Sound Termination on Programs with Infinite Models}

A different source of non-termination arises from programs with infinite three-valued models, as when the query {\tt p(?X)} is made to the following program. 
%
  \begin{tt}
    \begin{tabbing}
      foo\=foo\=f\=fooooo\=\kill
      ${\bf P_{InfModel}:}$ \\
      \> p(f(?X)) :- p(?X). \\
      \> p(a).
    \end{tabbing}
  \end{tt}
\noindent
Sound termination for WFS is possible using {\em restraint}
\cite{GroS13} a type of answer abstraction that sets the truth value
of abstracted answers to \undefined.  For instance, when using
restraint at term depth greater than or equal to two, the query {\tt
  p(?Y)} posed to $P_{InfModel}$ would produce {\tt p(a)} and {\tt p(f(a))}
both as \true answers, along with {\tt p(f(f(?X)))} with truth value
\undefined.  In this way, an informationally sound approximation to
the goal {\tt p(?Y)} is provided. Specifically, when using restraint
to evaluate a query to a program $P$, all atoms proved \true or \false
by \Ergo are \true or \false in the well-founded model of $P$.
However some atoms that are \true or \false in the well-founded model
may be \undefined in \Ergo.  Through restraint, \Ergo supports a fully
semantic approach to bounded rationality.

Answer abstraction can be requested with the help of the \texttt{answersize}  
restraint in the \texttt{setruntime} directive. Loading into \Ergo  the
small program below, 
\begin{verbatim}
   ?- setruntime{answersize(3,abstract)}.
   p(f(?X)) :- p(?X).
   p(a).
   ?- p(?X).  // query
\end{verbatim}
produces the following answers:
\begin{verbatim}
   ?X = a
   ?X = f(a)
   ?X = f(f(a))
   ?X = f(f(f(a)))
   ?X = f(f(f(?_h1945(?_h1946))))  - undefined
\end{verbatim}
Without the answer abstraction (i.e., without the \texttt{setruntime}
directive), the above query does not terminate.

\subsection{Tripwires}

From the viewpoint of \Ergo, both subgoal abstraction and restraint
based on answer abstraction are instances of {\em tripwires}.  A
tripwire allows a developer to specify that a property of an
evaluation should cause some intervention from the system.  The
intervention may be an automatic change in the resolution method, such
as with call abstraction or restraint; or the intervention could throw
an error; or suspend the computation for debugging.  \Ergo also
supports a {\em max answers} tripwire that allows a user to set a
limit on the number of answers a given subgoal $S$ may have.  If this
limit is reached, a new answer $A_S$ is added that is a variant of
$S$, the truth value of $A_S$ is set to \undefined and $S$ is
completed. Afterwards, queries to $S$ will return all of the derived
answers for $S$, as well as $A_S$.  The effect is to change the
closed-world assumption for $S$ from using \false as default to using
\undefined as default.  Another kind of tripwire is the ability to
time out a goal $G_{costly}$ that may either call \Ergo or an external
system.  The tripwire can be set either to add an answer $A$ that is a
variant of $G_{costly}$ with truth value \undefined, or to throw an
exception.  Finally, as mentioned in Section~\ref{sec:flexsearch}, a
negative subgoal that cannot be ground via delaying its execution can
be given the truth value \undefined instead of throwing an error.

{\revision
  In each of these types of tripwire a developer can weigh the benefit
of a bounded execution time against the possibility of deriving a set
of answers that are informationally sound but not complete.  It is
because of such uses that we denote the third WFS truth value as
\undefined, since it can mean {\em unknown} in addition to {\em
  undefined}.\footnote{In the case of procedural and other non-logical
modules, the action associated with a tripwire can be to throw an
error, or to call a user-defined handler.} }


\section{Architectural Summary}\label{sec:architecture}
\begin{centering}
\begin{figure}[hbtp]
\hspace{2cm}\includegraphics[trim={2cm 3cm 0cm 8cm},clip,scale=0.3]{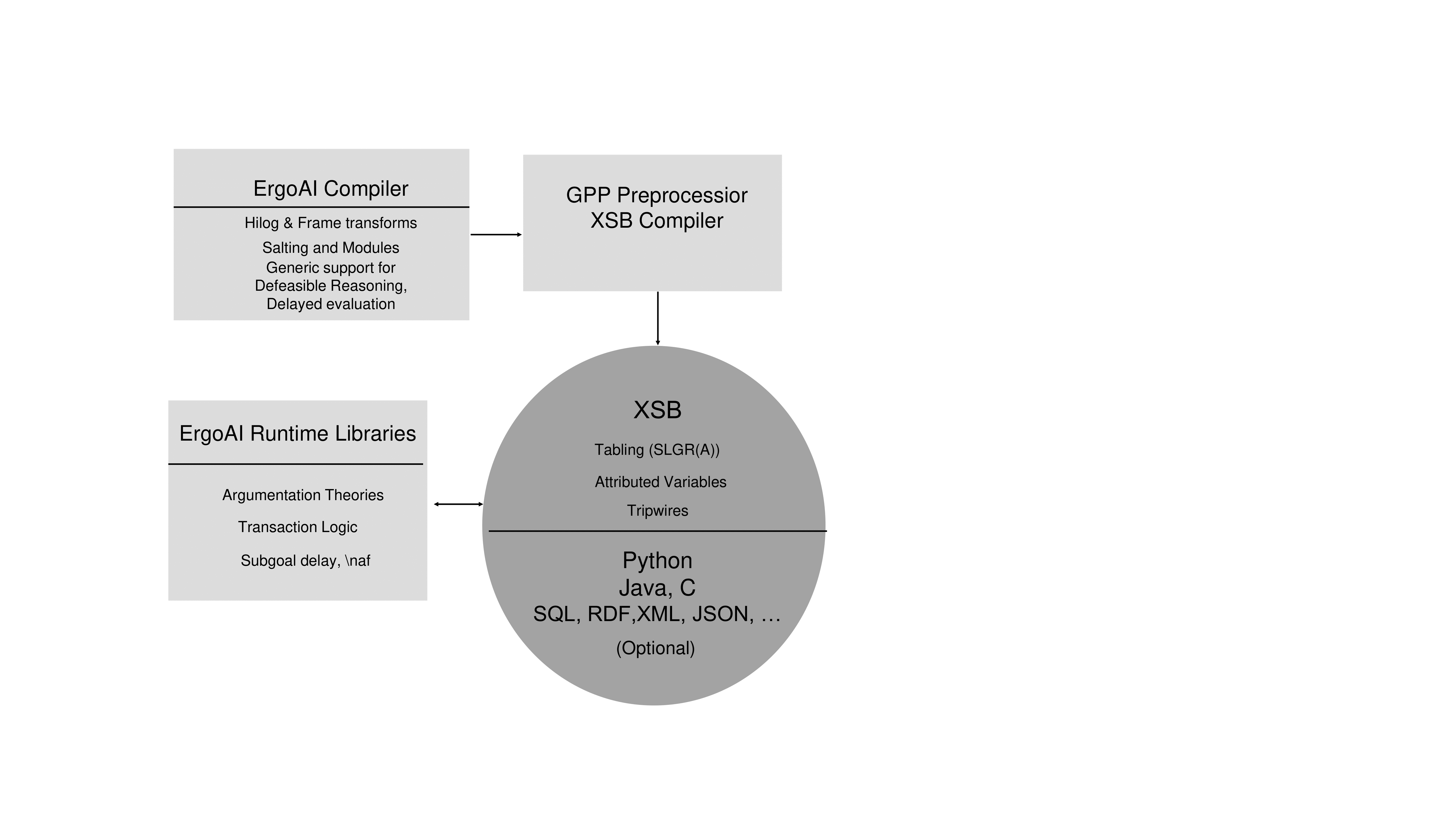}
  \caption{A schematic architecture for ErgoAI} \label{fig:arch}
\end{figure}
\end{centering}

Figure~\ref{fig:arch} illustrates the high-level architecture related to
the features presented in previous sections.  \Ergo is compiled into
tabled Prolog code that is executed by the XSB engine, along with
\Ergo's runtime libraries.  Optionally Janus
(Section~\ref{sec:python}) can load a Python engine into the same
process, leading to extremely efficient calls between the \Ergo and
Python. Direct interfaces to Java, C, SQL databases, SPARQL,
and direct connectors to XML, JSON, and RDF are also available.

More specifically, the \Ergo compiler performs the \hilog and frame
transformations of Section~\ref{sec:hilog} and compiles them
(Section~\ref{sec:frames}) first into GPP macros and then, at load time,
further into Prolog.
The first of these steps is performed by the
\Ergo compiler 
and the second by the GPP preprocessor.
Most of these macros provide a convenient way to specify how
module-dependent \emph{salting} is to be done.

Inheritance and other background theories, such as typing, are
implemented using \Ergo builders of GPP-enhanced Prolog as an
intermediate language.
In addition, the compiler performs transformations in support of
several aspects of defeasible reasoning
(Section~\ref{sec:argumentation}). This includes handling rule tags
and other rule descriptors as well as adding a defeasibility check to
each defeasible rule.  The resulting Prolog code makes use of {\em
  SLGR(A)} tabling (Section~\ref{sec:prelim}), attributed variables
for instantiation and \naf delay (Section~\ref{sec:flexsearch}) and
tripwires for bounded rationality (Section~\ref{sec:brat}).



       \section{Scalability and Performance} \label{sec:perf}
{\revision Because \Ergo is based on WFS it can be expected to be more
  scalable than KRR systems such as ASP or description logic solvers
  on transitive closure and other WFS-style queries.  Indeed, a recent
  paper benchmarks transitive-closure queries, comparing XSB's tabling both to
  the ASP solver Clingo and to the Datalog system Souffl\'e, showing
  excellent performance by tabling over a variety of
  graphs~\cite{LSIT25}.  One reason for this is that XSB, like \Ergo,
  does not require grounding, which is inefficient for large graphs.
  On the other hand, while ASP problems such as Hamiltonian path and
  graph coloring can be directly coded in \ErgoAI in a declarative
  manner, \ErgoAI's search strategy will perform worse than that of
  Clingo or other ASP systems.


Building on these results, we compare \Ergo's scalability and
performance to that of XSB's tabling.  We demonstrate that despite the
overhead of \hilog, frames, instantiation delay,
support for dynamic KRR, and other features, \Ergo remains within an
order of magnitude of tabled Prolog.\footnote{All benchmarks in this
section were performed on a 2023 MacBook Pro with an Apple M2 Pro chip
and 16 Gbyte RAM running Sonoma 14.7.2.}
We begin with a benchmark of Petri Net reachability that indicates the
general scalability of \Ergo.  We then separately benchmark
tabled and non-tabled programs. \ref{sec:perf-app} contains code for
all benchmarks.



\subsection{Performance Comparison for Petri Net Reachability}
Our first benchmark addresses the problem of reachability in 1-safe
Petri Nets, to which  the central problems of liveness, deadlock-freedom, and
the existence of home states can be reduced (cf. \cite{RosE98}).  The
benchmark, from \cite{MarS08b}, combines tabled and non-tabled
code,~\footnote{The XSB distribution includes a suite of Petri Net and
Workflow Net benchmarks some of which rely on constraint handling or
well-founded negation.}

A 1-safe Petri Net is a digraph of places and transitions: in
Figure~\ref{fig:Petri} the set of places is
\emph{\{p1,p2,b1,b2,c1,c2\}} while the set of transitions is
\emph{\{t1,t2,t3,t4\}}. Transitions are represented as facts such as:
\begin{alltt}
  \texttt{transition([b1,c1],[b2,c2],t1)}
\end{alltt}
indicating that if tokens are in places {\tt b1} and {\tt c1}, then
transition {\tt t1} can fire, placing tokens in {\tt b2} and {\tt c2}.
The state of a given Petri Net is represented as a list of the places
that contain a token: for example, in Figure~\ref{fig:Petri} the state
is represented as {\tt [b1,c1,p1]}.

\begin{figure}[htbp] 
  \centering \epsfig{file=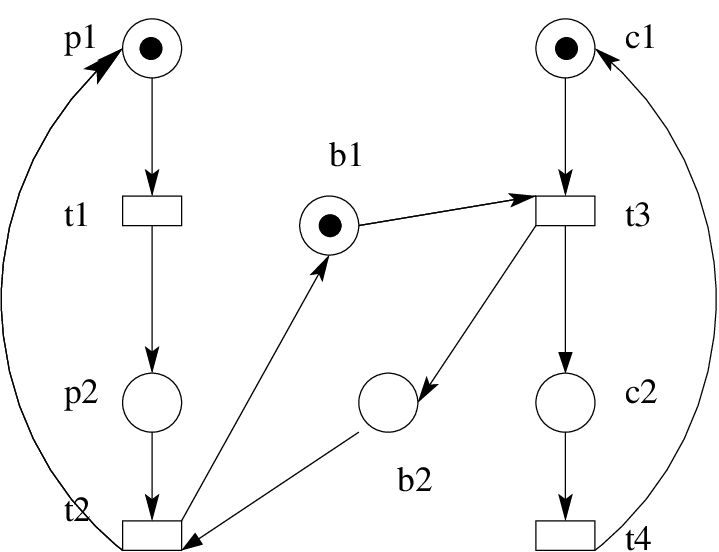,width=.4\textwidth}
  \caption{Illustration of a 1-safe Petri Net}
\label{fig:Petri}
\end{figure}

Figure~\ref{fig:petri-perf} compares \ErgoAI and XSB for reachability
queries to various sizes of generated Petri Nets.  Note that both the
time and space for \ErgoAI scale nearly linearly to a net with 8
million states, after which point RAM contention becomes an issue.
XSB's elapsed times are 4.7-6.0 times faster than \Ergo's, while
XSB's space is 4.0-4.7 times smaller than \ErgoAI's.

\begin{figure}[hbtp] 
 \begin{tabular}{rrrrrr} \\ 
   
  States        & 400,000 & 800,000  & 4,000,000  & 8,000,000 & 40,000,000
                                                                    \\ \hline \hline
 \Ergo Time     & 9.4 sec & 18.7 sec & 97.2 sec  & 202.0 sec   &  *  \\ 
 XSB  Time      & 1.7 sec &  3.1 sec & 19.5 sec  &  42.9 sec   & 244.0  \\\hline 
 \Ergo Memory   & 0.64 GB & 1.26 GB  & 6.22 GB   & 12.4 GB     &  *   \\ 
   XSB  Memory  & 0.16 GB & 0.31 GB  & 1.46 GB   &  2.9 GB     & 14.8    \\ 
 \end{tabular}
\caption{\ErgoAI/XSB elapsed time and space for Petri Net reachability}
\label{fig:petri-perf}
\end{figure}


\subsection{Performance Comparison for Tabled Resolution} \label{sec:tabled-perf}
           
Left-recursive transitive closure is a good benchmark for tabling,
since tabling performs nearly all computation apart from accessing the
graph edges.  Figure~\ref{fig:perf-tc} shows tests over a chain, a
cycle, and a chain in which all vertices have a self-loop.  For \Ergo,
measurements are performed for left-recursion over both \hilog
predicates and frames.  In these evaluations, left-recursion in frames
incurs a 20-40\% time overhead over predicates in \Ergo.  This is
because a frame query {\tt Id[Attr-$>$?Val]} must check for answers
through inheritance in addition to those obtained through transitive
closure.  For predicates, \Ergo's speed 
is consistently 3-4 times slower than XSB. The reason is that the
table space used by \Ergo is itself 3-4 times that of XSB due to
\Ergo's explanation structures.\footnote{This paper does not discuss
\Ergo's explanation mechanism which is in the process of being
redesigned and re-implemented. See Chapter 52 of the {\em \ErgoAI
  Reasoner User Manual} for a full description.}
CPU time scales linearly up to graphs with 100-200 million edges on
this platform.  Elapsed time has more variance, especially  when the memory needed
for table space by \Ergo becomes nearly the size of RAM (e.g., 14.5
Gbytes required out of 16 Gbytes available).

\begin{figure}[htp]\label{fig:perf-tc}
  \begin{scriptsize} 
    \begin{tabular}{lcrcrrrr} \\
  Graph   & Vertices   &HiLog  &Frame &XSB   & HiLog  &Frame&XSB \\
          &            & Time  &Time  &Time     &Space  &Space &Space
  \\ \hline \hline
chain     & $10^6$&0.21/0.23 &0.26/0.27  & 0.08/0.09 & 0.14 GB & 0.16 GB & 0.04 GB \\
chain     & $10^7$&1.95/2.08 &2.66/3.04  &0.57/0.64  & 1.4 GB  & 1.45 GB & 0.47 GB \\
chain     & $10^8$&20.1/34.9 &27.5/60.2  &5.43/5.90  & 14 GB   & 14.2 GB & 4.5 GB \\
\hline
self-loop &$10^6$& 0.39/0.40& 0.47/0.49  & 0.11/0.12 &0.14 GB& 0.16 GB & 0.04 GB \\
self-loop &$10^7$& 3.68/3.82& 4.85/5.03  & 0.94/0.98 &1.4 GB & 1.45 GB & 0.47 GB \\
self-loop &$10^8$& 37.4/49.4& 48.8/69.9  & 9.23/9.71 &14 GB  & 14.2 GB & 4.5  GB \\
\hline
cycle     &$10^6$& 0.21/0.23& 0.25/0.26 & 0.08/0.09 &0.14 GB& 1.65 GB   & 0.04 GB  \\
cycle     &$10^7$& 2.31/3.79& 2.61/2.74 & 0.58/0.64 &1.4 GB & 1.45 GB   & 0.44 GB  \\
cycle     &$10^8$& 23.6/40.0& 26.8/47.3 & 5.59/6.06 &14 GB  & 14.2 GB   & 4.5 GB 
\end{tabular}
\end{scriptsize}
\caption{CPU time/elapsed time in seconds, and space for
  left-recursive transitive closure}

\end{figure}  


  \subsection{Performance Comparison for Non-Tabled Resolution}
  General programming in \Ergo supports an open approach to knowledge
  formats, communicates with other components in a software stack,
  drives user interfaces, and enables much other functionality.  \Ergo
  is well-equipped to support general programming because it does not
  need to ground the entire program to evaluate a query and because it
  supports data structures through logical terms.  We use two simple
  predicates to benchmark general programming: {\tt makelist/2} to
  construct a list, and {\tt iter/2} to iterate through a loop
  (Figure~\ref{fig:makelist-iter}).  Both predicates perform
  structural recursion and do not need tabling.
  
  The \Ergo declaration {\em :- prolog\{f/n\}} causes a predicate to be
  compiled without tabling, although {\em f/n} is compiled with \Ergo
  features such as the ability to be disabled
  (Section~\ref{sec:dynamic}) and with two explanation arguments.
  We call such rules {\em quasi-Prolog rules}. As noted previously
  \Ergo can also call and be called from XSB Prolog, which provides
  another means to execute general programming tasks.

  Figures~\ref{fig:perf-makelist} and \ref{fig:perf-iter} compare
  these \Ergo execution methods with both static and dynamic Prolog code,
  where static code undergoes a deeper compilation process than
  dynamic code, but unlike dynamic code cannot be modified.
\begin{figure}[hbt] 
    \begin{tt}
      \begin{tabbing}
        fofo\=foofooooooooooooofoofooooooooooooo\=foofo\=\kill      
        makelist(0,[]):-!                    \> \>  iter(0):-!. \\
        makelist(?N,[?N|?Rest]):-            \> \>  iter(N):-  \\
        \> ?N1 \verb|\is| ?N-1,                    \> \> ?N1 \verb|\is| ?N-1, \\
        \> makelist(?N1,?Rest               \> \> iter(?N1).
      \end{tabbing}
    \end{tt}
    \caption{Test predicates to compare non-tabled resolution} \label{fig:nt-preds} \label{fig:makelist-iter}
\end{figure}
Figure~\ref{fig:perf-makelist} compares these execution methods on
{\tt makelist/2}.  All methods are able to scale to lists with ten
million elements. \Ergo calling static XSB code takes about the same
time as XSB calling static XSB code.  These methods are faster than
both \Ergo quasi-Prolog rules and XSB dynamic code.  However \Ergo
quasi-Prolog rules do not scale linearly for {\tt makelist/2}.
Analysis showed that non-linearity is due to the creation on
the heap of explanation structures that contain large lists.  Indeed,
when XSB's heap garbage collection is turned off, \Ergo quasi-Prolog
rules scale in a nearly linear manner.
As confirmation, Figure~\ref{fig:perf-iter} shows linear scaling for
quasi-Prolog rules on {\tt iter/1} which does not create large
structures.

  \begin{figure}[htp]
    \begin{sf}
    \begin{tabular}{lrrr}
Execution Method for {\tt makelist/1} & 100,000    & 1,000,000    & 10,000,000 \\ \hline \hline
\Ergo and XSB (static)   & $<0.01$    & 0.04      & 0.46 \\
\Ergo quasi-Prolog rules & 0.12       & 1.63          & 31.9\hspace{5pt} \\ 
XSB alone (static)      & $< 0.01$   & $0.02$        & 0.18  \\ 
XSB alone (dynamic)     & $0.02$     & 0.17          & 1.76 \\ \hline 
\Ergo quasi-Prolog rules (no heap gc)       & 0.07       & 0.74          & 10.0\hspace{5pt} \\*
XSB (dynamic, no heap gc)           & $< 0.01$   & 0.08          & 0.82 \\
    \end{tabular}
    \end{sf}
   \caption{Elapsed times in seconds to create lists of varying
     lengths using {\tt makelist/2}}\label{fig:perf-makelist}
  \end{figure}

  \begin{figure}[htp]
    \begin{sf}
    \begin{tabular}{lrrr}
Execution method for {\tt iter/1}    & 100,000    & 1,000,000    & 10,000,000 \\ \hline
\Ergo and XSB (static)  & $< 0.01$   & 0.02          & 0.14  \\
\Ergo quasi-Prolog rules  & 0.06       & 0.33          & 3.64\hspace{5pt} \\
XSB alone (static)            & $< 0.01$   & $0.02$        & 0.14  \\
\end{tabular}
\end{sf}
   \caption{Elapsed times in seconds for varying iterations using
     {\tt iter/1} in \Ergo and/or XSB}\label{fig:perf-iter}
  \end{figure}

  
  
  \subsection{Comparison of Internal Data Stores}
  \Ergo can access large factual knowledge bases in several ways, including:
\begin{itemize}
\item As full \hilog facts, loaded directly into \Ergo's internal
  knowledge store or accessed from external databases; or
\item As Prolog facts, loaded into XSB's internal knowledge store; or
\item As Prolog terms, accessed via Python connectors through Janus
\end{itemize}

As a scalability experiment, a KG of 11,847,275 triples from the DARPA
AIDA project (cf. Section~\ref{sec:defeas}) was loaded into the
knowledge stores of \Ergo and XSB.
\begin{itemize}
  \item {\em Loading the KG as \Ergo \hilog facts}.  Using
    \Ergo's \texttt{fastload} the KG was loaded at a rate of about 
    59,000 facts per second of elapsed time.  \Ergo's
    \texttt{fastload} uses trie indexing \cite{RRSSW98}.
  \item {\em Loading the KG directly into Prolog}. Using XSB's
    dynamic loading, the KG was loaded with full indexing on each fact
    (7 indexes) at a rate of about 137,000
    facts per second of elapsed time.
\end{itemize}
Of course, \Ergo also has its own interfaces for SQL, SPARQL, XML,
RDF, and other data sources.  For knowledge in virtually any other
format,\Ergo can access Python libraries through the high-speed Janus
interface (Section~\ref{sec:python}).

To summarize the overall performance and scalability results, \ErgoAI
is nearly as fast and scalable as tabled Prolog, while providing numerous
sophisticated KRR features.  For many applications these features are
well worth the cost in terms of expressiveness, ease of maintenance,
and updatability.  Put another way, \Ergo is a KRR system with
scalability and performance similar to a programming system.

%



\section{Discussion}\label{sec-conclusion}


\ErgoAI is a multi-paradigm logic programming system that makes
radically different choices from virtually all other logic programming
systems.  We have presented a few of these features, including
frames and inheritance for object-oriented logic programming based on
a solid semantics; \hilog predicates and terms
integrated into a powerful module system;
argumentation theories, the handling of unsafe queries, logical
approach to updates and management of change, along with support for
termination and bounded rationality.  We have not covered in detail
important features such as explanation; non-termination analysis;
user-defined functions; quantified logical formulas in rule bodies and
heads; typed variables and type constraints;
and many other features of the \Ergo language.

\Ergo is a significant system in itself, but also a testbed for
features that may be suitable for Prologs or other logic programming
paradigms.  For instance, the implementation of \hilog
(Section~\ref{sec:hilog}) may extend well to Prologs like SWI, whose
just-in-time indexing can adapt to the extra indexing burden required
by the \hilog and frame transforms
(Definition~\ref{def:hilog-transform}).  Transaction Logic
(Section~\ref{sec:dynamic}) with non-recursive update rules can be
implemented in Prolog via backtrackable dynamic code with hooks to
call integrity constraints.  Indeed the implementation of \Ergo has
shown that backtrackable updates can be implemented on top of regular
\texttt{assert/1} and \texttt{retract/1} quite efficiently.
However, a more complete implementation of
Transaction Logic requires tabling of the \emph{execution traces} of
transactions.
This remains difficult to efficiently, although \cite{FodK10} is a good step in this
direction.
The F-logic frames and classification hierarchies, whose compilation
into Prolog was outlined in Section~\ref{sec:frame-comp}
mainly requires compiler and reader extensions rather than engine
modifications, particularly in Prologs with strong
tabling and indexing.\footnote{Tabling is not strictly necessary for all aspects of
frames, but is highly desirable both to deal with recursion and to
efficiently implement inheritance.}

As mentioned in Section~\ref{sec:intro}, \Ergo has proven useful in
domains such as financial compliance, tax accounting, and legal
reasoning. More recently, in the DARPA CODORD project, \Ergo code
is generated from natural language narratives and policy documents by
five different automated systems based on LLMs, ILPs, semantic parsing
and other techniques.  Our near-term work includes continued
improvements in \Ergo's speed and robustness, as well as improvements
in explanation generation.  Longer-term objectives include
supporting fuzzy and probabilistic reasoning techniques for \hilog
predicates.  The approach under design uses techniques from the
PITA~\cite{RigS11a} system.  PITA, which was originally implemented
for XSB, is based on a source-level transformation followed by
execution that uses tabling and BDDs.  An initial extension to \Ergo's
\hilog predicates requires reimplementing PITA's source-code
transformation within the \Ergo compiler.

\textbf{Acknowledgments}.
We are grateful to Benjamin Grosof and Paul Fodor for many technical
and visionary discussions. Many of the additions to \Ergo were
inspired by this collaboration.
We thank Daniel Elenius for suggesting an idea that gave rise to the
\texttt{RefuteCLP} argumentation theory discussed in
Section~\ref{sec:defeas}.
We are also grateful to Martin Gebser and Torsten Schaub for their
comments on the suitability of various problems for ASP and tabled
Prolog.



\bibliographystyle{acmtrans}
\bibliography{longstring,all,main}

\newpage
\appendix
\section{Code Used for Benchmarks in Section~\ref{sec:perf}} \label{sec:perf-app}

\subsection{Code Used to Benchmark 1-Safe Petri Nets} \label{sec:petri}
  
\begin{figure}[htbp]
\begin{verbatim}
:- table reachable/2.
reachable(InState,NewState):-
   reachable(InState,State),
   hasTransition(State,NewState).
reachable(InState,NewState):-
   hasTransition(InState,NewState).

hasTransition(State,NewState):-
   get_trans_for_conf(State,AllTrans),
   member(Trans,AllTrans),
   apply_trans_to_conf(Trans,State,NewState).

get_trans_for_conf(State,Flattrans):-
   get_trans_for_conf_1(State,State,Trans),
   flatten(Trans,Flattrans).

get_trans_for_conf_1([],_State,[]).
get_trans_for_conf_1([H|T],State,[Trans1|RT]):-
   findall(trans([H|In],Out,Tran),trans([H|In],Out,Tran),Trans),
   check_concession(Trans,State,Trans1),
   get_trans_for_conf_1(T,State,RT).

check_concession([],_,[]).
check_concession([trans(In,Out,Name)|T],Input,[trans(In,Out,Name)|T1]):-
   ord_subset(In,Input), 
   ord_disjoint(Out,Input),!,
   check_concession(T,Input,T1).
check_concession([_Trans|T],Input,T1):-
   check_concession(T,Input,T1).

apply_trans_to_conf(trans(In,Out,_Name),State,NewState):-
   ord_subtract(State,In,Diff),
   flatten([Out|Diff],Temp),
   sort(Temp,NewState).
\end{verbatim}
\caption{Tabled Prolog program for reachability in a 1-safe Petri Net}
\label{fig:petri-code-XSB}
\end{figure}

The tabled Prolog program for Petri Net reachability
(Figure~\ref{fig:petri-code-XSB}) was introduced in \cite{MarS08b} and
is available in the XSB distribution under the path {\tt
  mttests/benches/petri}. Both the Prolog and the \Ergo version
(Figure~\ref{fig:petri-code-ergo}) perform left-recursive transitive
closure in which the base case backtracks through states to which
there is a direct transition from the input state.  Because a state of
a Petri Net is represented as a list of the places that contain a
token, determining the new states requires a small amount of list
manipulation.
The \Ergo version differs in small ways from the tabled Prolog
version. In particular, predicates that are not to be tabled (called
{\em transactional predicates}) have names that begin with \verb|%|.
Because there is a small overhead to transactional predicates in
\Ergo, some predicate unfolding has been done in \Ergo compared to
the tabled Prolog version.  Note, however that XSB list library
predicates such as {\tt member/2}, {\tt flatten/3}, and {\tt sort/2}
are called directly, using module qualification.  Also note that
aggregation predicates like Prolog's {\tt setof/3} have a different
syntax in \Ergo.

\begin{figure}[htbp]
\begin{verbatim}
reachable(?InState,?NewState):-
   reachable(?InState,?State),
   hasTransition(?State,?NewState).
reachable(?InState,?NewState):- hasTransition(?InState,?NewState).

hasTransition(?State,?NewState):-
   %get_trans_for_state_1(?State,?State,?MidTrans),
   flatten(?MidTrans,?AllTrans)@\plg(basics),
   member(trans(?Input,?Out,?_Name),?AllTrans)@\plg(basics),
   ord_subtract(?State,?Input,?Diff)@\plg(ordsets),
   flatten([?Out|?Diff],?Mid1)@\plg(basics),
   sort(?Mid1,?NewState)@\plg.

%get_rules_for_state_1([],?_State,[]):- \true.
%get_rules_for_state_1([?H|?T],?State,[?Trans1|?TransTail]):-
   ?TransSet = setof{?Trans | trans([?H|?Places],?Output,?Tran)@\plg(gen_elem),
                              ?Trans = trans([?H|?Places],?Output,?Tran) },
   %check_concession(?TransSet,?State,?Trans1),
   %get_rules_for_state_1(?T,?State,?TransTail).
	
// Check that a token is in all places in *t, and no token is in t*
%check_concession([],?_,[]):- \true.
%check_concession([rule([?_Inp|?Ilist],?Outlist,?_Name)|?T],?Input,
                  [rule([?_Inp|?Ilist],?Outlist,?_Name)|?T1]):-
   ord_subset(?Ilist,?Input)@\plg(ordsets),
   ord_disjoint(?Outlist,?Input)@\plg(ordsets),!,
   %check_concession(?T,?Input,?T1).
%check_concession([?_Rule|?T],?Input,?T1):-
   %check_concession(?T,?Input,?T1).
\end{verbatim}
\caption{\Ergo program for reachability in a 1-safe Petri Net}
\label{fig:petri-code-ergo}
\end{figure}


\subsection{Code Used for Benchmarking Tabled Resolution}

\begin{figure}[hbt]
\begin{verbatim}
tc_cycle(?Limit,?From,?To):- 
    edge_cycle(?Limit,?From,?To)@\plg(tc_cycle).
tc_cycle(?Limit,?From,?To):-
    tc(?Limit,?From,?Mid),
    edge_cycle(?Limit,?Mid,?To)@\plg(tc_cycle).
\end{verbatim}
\caption{\Ergo code for transitive closure over large cycles}
\label{fig:ergo-cycle}
\end{figure}

Figures~\ref{fig:ergo-cycle} and \ref{fig:prolog-cycle} show the code
used in benchmarks for transitive closure.  Note that the \Ergo
transitive closure code calls the Prolog {\tt edge/2} predicate
(Section~\ref{sec:tabled-perf}). The code is shown for cycles only;
code for chains and chains with self-loops is similar.

\begin{figure}[hbt]
\begin{verbatim}
:- export tc_cycle/3, edge_cycle/3.

:- table tc_cycle/3.
tc_cycle(Limit,From,To):- edge_cycle(Limit,From,To).
tc_cycle(Limit,From,To):-
    tc_cycle(Limit,From,Mid),
    edge_cycle(Limit,Mid,To).

edge_cycle(Limit,From,To):-
    (Limit >= From ->
         (To is From+1 ; To=From)
      ;  fail).
\end{verbatim}
\caption{Prolog code for transitive closure over large cycles}
\label{fig:prolog-cycle}
\end{figure}

\subsection{Code Used for Benchmarking Non-Tabled Resolution} \label{sec:app-nontabled}

\paragraph{\Ergo quasi-prolog rules.}

In a module for which production mode is enabled, \Ergo quasi-prolog
rules are declared via {\tt ?- prolog\{/1\}},
which causes compilation to omit the \hilog transformation, and
ensures that the rule is not tabled.  For these experiments, subgoal
delay was manually turned off for these rules~\footnote{Automatic
subgoal delay is not needed for Prolog predicates, and can be
expensive for large terms, such as those created in {\tt makelist/2}.}
but extra arguments used for explanation, are added to the compiled
version of the predicates.

\begin{figure}[hbt]
    \begin{tt}
      \begin{tabbing}
        f\=foo\=ooooooooooofoofooooooooooooooooooooooo\=foo\=\kill      
 \> :- compiler\_options\{production=on\}. \\
\\
\> ?- disablefeature\{subgoal\_delay\}.\\

 \> :- prolog\{makelist/2\}.     \>\>           :- prolog\{iter/1\}.\\
\> makelist(0,[]):- !.          \>\>           iter(0):-!. \\  
\> makelist(?N,[?N|?T]):-       \>\>           iter(?N):- \\              
\> \> 	is(?N1 , ?N - 1)@\plg,  \>\>              is(?N1 , ?N - 1)@\plg,       \\ 
\> \> 	makelist(?N1,?T).       \>\>              iter(?N1).
      \end{tabbing}
    \end{tt}
    \caption{\Ergo quasi-prolog rules used for benchmarking}
\end{figure}

  \paragraph{\Ergo rules that call XSB.}
  Calling XSB from \Ergo is shown below.
  
    \begin{tt}
      \begin{tabbing}
        f\=foo\=ooooooooooofoofooooooooooooooooooooooo\=foo\=\kill      
        \> makelist(?N,?L):-                            \>\>   iter(?N):- \\        
        \> \> 	plg\_makelist(?N,?L)@\plg(makelist).    \>\>     plg\_iter(?N)@\plg(iter).
      \end{tabbing}
    \end{tt}


\end{document}